# QUARK GLUON PLASMA IN NUMERICAL SIMULATIONS OF LATTICE QCD

CARLETON DeTAR

*Physics Department, University of Utah*
*Salt Lake City, Utah 84112, USA*
E-mail: detar@physics.utah.edu

## ABSTRACT

Numerical simulations of quantum chromodynamics at nonzero temperature provide information from first principles about the physical properties of the quark gluon plasma. Because the lattice approximation can be refined indefinitely, results of lattice simulations now provide the most reliable basis for our understanding of the nonperturbative characteristics of the plasma and of the high temperature phase transition. Following a brief overview of the methodology of lattice gauge theory at nonzero temperature, recent results and insights from lattice simulations are discussed. These include our understanding of the phase diagram of QCD, the nature of the phase transition, and the structure of the plasma.

## Contents

hep-ph/9504325  18 Apr 1995

**1. Introduction**

In 1974 Wilson proposed a formulation of quantum chromodynamics (QCD) on a discrete space-time lattice.[1] A few years later Creutz demonstrated that this formulation was a promising basis for the successful numerical simulation of the theory.[2] The simulation method is suitable for studies of QCD at nonzero temperature in *thermal equilibrium* and, at least at present, at *zero baryon chemical potential*. It is the most promising currently available method for deducing nonperturbative characteristics of QCD at zero or nonzero temperature, starting from first principles. Given suitable algorithms and computing power, the only approximations, namely, finite lattice spacing, finite volume, and unphysically large quark masses, can be refined indefinitely. Thus lattice simulations provide the basis for much of our theoretical understanding of the characteristics of QCD at nonzero temperature.

Are the predictions of lattice QCD relevant to the experimental effort? The violently expanding product of a high energy heavy ion collision can scarcely be approximated as an equilibrium plasma at zero baryon chemical potential. Indeed, even a hydrodynamical description that assumes local equilibrium could be questionable. Thus despite our confidence in lattice QCD, it is still essential that we formulate good models to bridge the gap between the relatively solid predictions of the computer simulations and the experimental environment.

In the past few years lattice simulations have achieved a reasonably consistent qualitative picture of the high temperature behavior of QCD. What has emerged is a far more subtle and intriguing picture of the quark gluon plasma and the nature of the phase transition than earlier caricatures would have suggested. At very high temperatures we have a plasma that behaves in bulk roughly like a free gas of quarks and gluons, yet it retains features of confinement that are evident in long-range correlations. Although there may be no phase transition, there is at least a dramatic crossover accompanied by dramatic changes in the structure and symmetries of the plasma. In this chapter we will be taking stock of what lattice QCD tells us about hadronic matter in thermal equilibrium. Particular emphasis will be placed on developments since the last volume, which featured an excellent review by Karsch.[3] For the past several years, the annual "Lattice" conference has regularly featured sessions on nonzero temperature QCD. Recent proceedings can be found in Refs. 4,5,6,7.

## 2. Nonzero Temperature QCD on the Lattice

Here we give a very brief sketch of the basis for lattice gauge theory at zero and nonzero temperature. Our purpose is to indicate how the Euclidean space-time lattice with periodic boundary conditions arises naturally in the quantum Gibbs ensemble, and to show how the temperature of the ensemble determines the imaginary time dimension of the lattice. For a more complete exposition, there are some excellent texts.[8]

*2.1. Feynman Path Integral from the Gibbs Ensemble: A Thumbnail Sketch*

Nonzero temperature simulations of a field theory are designed to calculate operator expectation values on the quantum Gibbs ensemble at temperature $T$:

$$\langle \mathcal{O} \rangle = \mathrm{Tr}(\mathcal{O} e^{-H/T}) / \mathrm{Tr} e^{-H/T}, \tag{1}$$

where $H$ is the hamiltonian for the field theory, $\mathcal{O}$ is any operator, the trace is over all physical states in the Hilbert space, and we use units in which the Boltzmann constant $k$ is one. The operator $\exp(-H/T)$ is just the standard time evolution operator $\exp(-iHt)$ evaluated at an imaginary time

$$t = -i/T. \tag{2}$$

This expression is the origin of the imaginary time coordinate and the important relationship between the temperature and the extent of the Euclidean spacetime volume in imaginary time.

Several steps are required in order to convert the trace over states in the Gibbs ensemble (1) into a multidimensional integral over lattice variables. Let us consider briefly how this is done for a scalar field theory based on a single field $\phi$. First the hamiltonian $H$ is formulated on a three-dimensional lattice with lattice constant $a$. The continuous field then takes on values $\phi(\mathbf{x})$ on each of the sites $\mathbf{x}$ of a three-dimensional lattice. The trace over states is written on a complete orthonormal basis

$|\phi(\mathbf{x})\rangle$ in which the field is diagonal. To facilitate the estimation of the time evolution operator, the imaginary time interval $[0, 1/T]$ is then subdivided into $N_t$ steps for large $N_t$, and the trace (partition function) is rewritten in the form

$$Z = \text{Tr} e^{-H/T} = \text{Tr}(e^{-H/N_tT} e^{-H/N_tT} \ldots e^{-H/N_tT}). \tag{3}$$

This step produces the lattice discretization in imaginary time. Each of the infinitesimal time evolution operators (transfer matrices) in the product is written as a matrix on the field-diagonal basis. This is done through the completeness relation

$$1 = \int \prod_{\mathbf{x}} d\phi(\mathbf{x}) |\phi(\mathbf{x})\rangle\langle\phi(\mathbf{x})|. \tag{4}$$

Since this relation is inserted between each factor, it is convenient to introduce an extra label $\tau$ to distinguish the multiplte integration variables: $\phi(\mathbf{x}, \tau)$. The extra variable is naturally taken to be a discrete imaginary time variable, leading to a classical field variable $\phi(\mathbf{x}, \tau)$ defined on a four-dimensional Euclidean lattice. The discrete time values are

$$\tau = a_t k \quad \text{for } k = 0, 1, 2, \ldots, N_t - 1. \tag{5}$$

where $a_t = 1/N_tT$ is taken to be the lattice constant in the imaginary time direction. In terms of this labeling the partition function then becomes

$$Z = \int \prod_{\mathbf{x},\tau} d\phi(\mathbf{x}, \tau) \langle\phi(\mathbf{x}, 0)| e^{-H/N_tT} |\phi(\mathbf{x}, (N_t - 1)a_t)\rangle \tag{6}$$

$$\langle\phi(\mathbf{x}, (N_t - 1)a_t)| e^{-H/N_tT} |\phi(\mathbf{x}, (N_t - 2)a_t)\rangle \ldots$$
$$\langle\phi(\mathbf{x}, 2a_t)| e^{-H/N_tT} |\phi(\mathbf{x}, a_t)\rangle \langle\phi(\mathbf{x}, a_t)| e^{-H/N_tT} |\phi(\mathbf{x}, 0)\rangle.$$

Since we are taking a trace, we have built in the requirement that $\phi(\mathbf{x}, N_t a_t) = \phi(\mathbf{x}, 0)$. This is the origin of periodicity in imaginary time. An explicit evaluation of the transfer matrix elements leads to Feynman's remarkable path integral formula for the quantum partition function

$$Z(T) = \text{Tr} e^{-H/T} = \int \prod_x d\phi(x) \exp[-S(\phi, T)], \tag{7}$$

where $S(\phi, T)$ is the imaginary time classical action for the field configuration $\phi(x)$ for $x = (\mathbf{x}, \tau)$ on the Euclidean space-time lattice of dimension $N^3 \times N_t$. The integration is over all possible choices of the field values on the lattice.

Any observable $\mathcal{O}$ is a function of the field $\phi$. The expectation value (1) is then

$$\langle\mathcal{O}\rangle = \int \prod_x d\phi(x) \mathcal{O}(\phi) \exp[-S(\phi, T)]/Z(T) \tag{8}$$

## 2.2. Monte Carlo Methods

As sketched in the previous section, the Feynman path integral formulation reduces quantum statistical mechanics to an integration over classical variables. Formulated on a lattice, the problem is reduced to a multidimensional integration. For a wide class of actions, the weight factor $\exp(-S)$ of the integration is positive definite, and the integration can be done effectively using a variety of Monte Carlo sampling methods, using the weight factor as a probability. Among these are heat bath, Metropolis, and molecular dynamics methods. The basic idea is to produce a large biased sample of points $\{\phi_i(x), \ i = 1, \ldots, N_{\text{conf}}\}$ in the space of the multidimensional integration. These are commonly called "configurations", and are characterised by choosing a particular value of the field on each space-time point. The sample is biased so that the probability of encountering a configuration is proportional to the weight factor

$$P(\{\phi(x)\}) \propto \exp(-S). \tag{9}$$

On such a biased sample, the operator expectation value is simply the average of its values on the sample configurations:

$$\langle \mathcal{O} \rangle = \lim_{N_{\text{conf}} \to \infty} \frac{1}{N_{\text{conf}}} \sum_{i=1}^{N_{\text{conf}}} \mathcal{O}(\phi_i). \tag{10}$$

The algorithmic challenge lies in formulating an efficient sampling method that produces a desired variance at the lowest computational cost.

## 2.3. Lattice QCD

### 2.3.1. Pure Yang-Mills Theory

The Feynman path integral is expressed in terms of the action in a Euclidean space time. The Wilson formulation of the Euclidean action for quantum chromodynamics starts from a regular hypercubic lattice with equal space and time lattice constants $a = a_t$. The gauge vector potential $A_\mu^a(x)$ defines the gauge connection

$$U_{x,\mu} = \exp(igaA_\mu^c(x)\lambda_c/2) \tag{11}$$

between the site $x$ and the nearest neighbor site $x + \hat{\mu}$. (The lattice vector of length $a$ in the $\mu$ direction is $\hat{\mu}$.) Here $g$ is the gauge coupling constant and $\lambda_c$ are the usual generators of the SU(3) Lie algebra. This SU(3) matrix is called the gauge link matrix. There is one such variable for each of the links connecting nearest neighbors on the lattice. A forward connection for a given link is associated with the matrix $U$ and a backward connection for the same link is associated with the adjoint of that matrix $U^\dagger$.

The plaquette variable is defined on a unit square on the lattice as the product of the connections around the square.

$$\square = \text{Tr} U_{\mu\nu}(x) = \text{Tr} U_\mu(x) U_\nu(x+\hat{\mu}) U_\mu^\dagger(x+\hat{\nu}) U_\nu^\dagger(x) \qquad (12)$$

The trace of any such product of gauge connections around a closed path is gauge invariant. The plaquette variable is related to the SU(3) color Maxwell field strength in the continuum limit.

$$\lim_{a \to 0} \text{ReTr} U_{\mu\nu}(x)/3 = 1 - \frac{a^4 g^2}{6}[F_{\mu\nu}^c(x)]^2 \qquad (13)$$

The continuum Euclidean action for a pure gluon field is

$$S_g = \int d^4x \frac{1}{4}[F_{\mu\nu}^c(x)]^2. \qquad (14)$$

Wilson suggested the lattice approximation

$$S_g = \sum_x \sum_{\mu \neq \nu} 6/g^2 [\text{ReTr} U_{\mu\nu}(x)/3 - 1] \qquad (15)$$

Creutz first demonstrated the feasibility of using this simple approximation in numerical simulations. It has served as the basis for a great many studies since then. Improvements, some very promising, have been proposed. They add terms to the action formed from products of the gauge connection around larger loops.[9,10,11] The objective is to remove lattice artifacts as nearly as possible and facilitate the approach to the continuum limit.

### 2.3.2.Including Quarks

Incorporating fermions into the functional integral demands an additional effort. The Pauli exclusion principle requires that they be introduced as anticommuting Grassmann numbers rather than the ordinary commuting numbers of the boson fields. Since we compute with ordinary numbers, it is then necessary to complete the integration over the fermion degrees of freedom by hand. Fortunately, this is easy to do. Unfortunately, the resulting nonlocal effective action vastly increases the computational cost. The result, however, is a simulation of full QCD.

An additional complication with fermions is a difficulty in controlling the number of fermion species. We would like to construct a theory that reproduces faithfully the chiral symmetry of the continuum theory at zero quark mass. The lattice regularization forces us to make a difficult choice. Either we give up chiral symmetry or we must have a doubling (usually a few redoublings) of the number of quark species.

There are two popular lattice fermion formulations corresponding to these choices. One is called the staggered or Kogut-Susskind fermion formulation[12] and the other the Wilson fermion formulation.[1] The hope is that in the continuum limit, either choice takes us to the one and only continuum theory.

For present purposes we merely write down the lattice fermion actions for these two choices. For further details the reader should consult Ref. 8. In the Wilson fermion formulation the quark field for each flavor is represented as a Dirac color spinor $\psi_j^c(x)$ on each lattice site $x$ with a three-component color index $c$ and a four-component Dirac spin index $j$. The fermion action is then

$$S_f^W = \sum_x \left\{ \bar\psi(x)\psi(x) - \kappa \sum_{\mu=1}^4 [\bar\psi(x+\hat\mu)(r+\gamma_\mu)\psi(x) + \bar\psi(x)(r-\gamma_\mu)\psi(x+\hat\mu)] \right\} \quad (16)$$

where $\kappa = 1/(2am + 8r)$ and $r$ is usually taken to be 1. The summation over color and spin degrees of freedom is implicit. With $r = 0$ the fermion action is chirally symmetric at zero quark mass and describes 16 degenerate fermion species. With $r \neq 0$ the degeneracy is lifted at the expense of destroying chiral symmetry in the zero quark mass limit.

In the "staggered" fermion formulation the quark field is represented as a color spinor $\psi^c(x)$ with no explicit Dirac spin degree of freedom. In effect, four spin and four flavor components are distributed over each hypercube of dimension $2^4$. The fermion action is

$$S_f^S = \sum_x \left\{ 2am\bar\psi(x)\psi(x) + \sum_{\mu=1}^4 \alpha_\mu(x)[\bar\psi(x)\psi(x+\hat\mu) - \bar\psi(x+\hat\mu)\psi(x)] \right\} \quad (17)$$

The phase factors $\alpha_\mu(x)$ are diagonalized Dirac matrices. The summation over the color index is implicit. The theory is chirally symmetric at zero quark mass, but there are four degenerate flavors. Such a flavor symmetry is unnatural, but there are methods for reducing the effective flavor number.[8] Because flavor rotations involve fields on different lattice sites, except for U(1) transformations, which are diagonal in space, flavor rotations are restricted to a discrete lattice subgroup. Thus instead of the full SU(4)×SU(4) chiral symmetry, the symmetry consists of a U(1)×U(1) subgroup plus a discrete subgroup. Since a restoration of rotational invariance in the continuum limit is anticipated, it is also expected that the full chiral symmetry will be restored simultaneously. On a coarse lattice members of the same flavor multiplet are not necessarily degenerate. For example, it is popular to measure the masses of two members of the pion multiplet, often called $\pi$ and $\pi_2$, with local operators in the staggered fermion scheme. The mass ratio $m_\pi/m_{\pi_2}$ serves as an indicator of progress toward the continuum limit, since it should approach one.

The total action in either case is the sum of the gauge and fermion parts

$$S = S_g + S_f. \quad (18)$$

The partition function is then given by

$$Z = \int dU d\psi d\bar\psi e^{-S}. \tag{19}$$

As we have noted, the fermion fields must be integrated out explicitly. In either formulation, the fermion fields enter in a bilinear form,

$$S_f = \bar\psi M(U)\psi \tag{20}$$

where the fermion matrix $M(U)$ is a matrix with row and column indices labeled by the spatial coordinate as well as color and, if necessary, spin. Integrating out the quark degrees of freedom leads to

$$\int \prod_x d\bar\psi(x) d\psi(x) \exp[\bar\psi M(U)\psi] = \det M(U) \tag{21}$$

Thus the effective gauge action is

$$S_{\text{eff}}(U) = S_g(U) + \log \det M(U) \tag{22}$$

and the partition function is then

$$Z = \int dU e^{-S_{\text{eff}}(U)} \tag{23}$$

The second term in the effective action is the fermion determinant. It depends on the gauge fields and induces a nonlocal gauge field interaction. This feature vastly increases the computational effort. It is beyond the scope of this review to discuss the various methods for accommodating the fermion determinant in a numerical simulation.[8] The most effective method uses a molecular dynamics approach. With tricks it is possible to approximate the induced effect upon the gauge field of any number of fermion species. The more elegant "exact" simulations require four species of staggered fermions or two species of Wilson fermions.

### 2.3.3. Asymptotic Freedom

Since our goal in lattice QCD is to regulate and approximate the continuum theory, it is crucial that there be a meaningful lattice continuum limit. Fortunately, QCD (and lattice QCD) is an asymptotically free field theory with an ultraviolet fixed point. As the lattice coupling $g$ is decreased, the lattice constant $a$, measured in physical units, also decreases. For small enough $g$, it decreases according to the perturbative scaling relation

$$a\Lambda = e^{-1/(2\beta_0 g^2)}(\beta_0 g^2)^{-\beta_1/(2\beta_0^2)} \tag{24}$$

where $\beta_0 = (11-2N_f/3)/(16\pi^2)$ and $\beta_1 = (102-38N_f/3)/(16\pi^2)^2$ and $\Lambda$ sets the scale. It is determined in lattice regularization from a physical quantity, such as the proton mass. A proton may occupy only a few lattice sites at large $g$, but as $g$ is decreased, the proton occupies more lattice sites, so is represented with increasing resolution. It follows from the relationship between $g$ and lattice scale $a$, that increasing $6/g^2$ corresponds to increasing the temperature $T = 1/(N_t a)$. Thus by varying $6/g^2$, it is possible to study a range of temperatures without changing the number of lattice points.

### 2.3.4. Quenched Approximation

The "quenched" approximation to QCD amounts to carrying out the simulation with the fermion determinant set to one, cutting the computational effort by orders of magnitude. In this case the thermal ensemble consists only of gluons. Quite useful results can be obtained in this way at zero temperature, where it has been difficult to find cases where vacuum fermion loops seem to make a difference. At nonzero temperature, however, fermions do make a significant difference in the thermal ensemble. The phase structure changes when quarks are included. Thus the quenched approximation at nonzero temperature may give suggestive results, but fermions cannot be neglected if precise contact with nature is needed.

### 2.3.5. Nonzero Chemical Potential

The partition function for full QCD, Eq. (19), represents the grand canonical ensemble at zero baryon chemical potential. The fermion determinant is real and nonnegative for two flavors of Wilson fermions or four flavors of staggered fermions. Thus the factor $\exp(-S_{\text{eff}})$ is a suitable probability weight for a Monte Carlo simulation. It would be very useful to be able to carry out simulations at nonzero chemical potential. The problem is obviously important, since the debris of a heavy ion collision necessarily includes regions of nonzero chemical potential and the cores of neutron stars are obviously baryon rich. Unfortunately at nonzero chemical potential, we encounter a fundamental technical problem: The fermion determinant becomes complex, spoiling the probability weight. Various tricks have been proposed to evade this problem. They include carrying out simulations in the quenched approximation or at zero chemical potential, and incorporating a correction factor to simulate the effects of a nonzero chemical potential. Such methods are limited to a very small chemical potential or a very small volume. Thus they may succeed in giving hints about what happens at a small chemical potential, but none has been successful in achieving a result that can be confidently extended to the thermodynamic limit of inifinite volume. Thus this problem remains an open challenge. For a recent review, see Ref. 13.

*2.4. Lattice Observables*

### *2.4.1. Quark Propagator*

To determine properties of the thermal ensemble one measures a variety of observables. Many of the important observables involve quarks. Thus, for example, we need the quark propagator

$$\langle \psi(z)\bar\psi(y) \rangle = Z^{-1} \int \prod_{x,\mu} dU_\mu(x) d\psi(x) d\bar\psi(x) e^{-S} \psi(z)\bar\psi(y) \tag{25}$$

The integration over quark variables is again carried out explicitly, giving

$$\langle \psi(z)\bar\psi(y) \rangle = Z^{-1} \int \prod_{x,\mu} dU_\mu(x) M^{-1}(z,y,U) \exp(-S_{\text{eff}}(U)) \tag{26}$$

or the gauge field average of the inverse of the fermion matrix. (We have suppressed the color and spin labels for simplicity.) In a more compact notation, we may write

$$\langle \psi(z)\bar\psi(y) \rangle = \langle M^{-1}(z,y,U) \rangle_U \tag{27}$$

where the subscript $U$ in the expectation value on the right side indicates that it is taken with respect to the effective gauge action.

### *2.4.2. Hadron Propagator*

If an observable is not gauge invariant, the integration over gauge variables gives zero. The quark propagator we have been discussing can be defined in a specific gauge. A hadron propagator, on the other hand, is gauge invariant. For example the operator $\bar\psi(x)\Gamma\psi(x)$ is an interpolating field for a quark-antiquark meson with a particular Dirac matrix $\Gamma$ determining the spin and parity. Thus a meson propagator can be extracted from the correlation

$$\begin{aligned}
& \langle \bar\psi(z)\Gamma\psi(z)\bar\psi(y)\Gamma\psi(y) \rangle_U - \langle \bar\psi(0)\Gamma\psi(0) \rangle_U \langle \bar\psi(0)\Gamma\psi(0) \rangle_U \\
= & \langle \text{Tr}[M^{-1}(z,y,U)\Gamma M^{-1}(y,z,U)\Gamma] \rangle_U \\
- & \langle \text{Tr}[M^{-1}(z,z,U)\Gamma] \text{Tr}[M^{-1}(y,y,U)\Gamma] \rangle_U \\
- & \langle \text{Tr}[M^{-1}(0,0,U)\Gamma] \rangle_U \langle \text{Tr}[M^{-1}(0,0,U)\Gamma] \rangle_U
\end{aligned} \tag{28}$$

The second term on the right side makes a contribution only for flavor singlet mesons. It represents a coupling to gluon intermediate states. The last term, the vacuum disconnected term, on either side contributes only for flavor singlet mesons with vacuum quantum numbers, such as the chiral condensate order parameter $\bar\psi\psi$.

In this way all hadron propagators are constructed from products of the inverse of the Dirac matrix, averaged over gauge field configurations with a weight determined by the effective gauge action.

### 2.4.3. Polyakov Loop

An important observable simulates the effect of introducing a static spinless quark into the ensemble. It is formed from the product of a string of gauge link matrices along a line in the time direction. The observable that introduces a static quark at site **x** is

$$P(\mathbf{x}) = \text{Tr} \prod_{\tau=0}^{N_t-1} U_4(\mathbf{x},\tau) \qquad (29)$$

where the trace is over the color degrees of freedom. The static quark world line is closed by virtue of the periodicity of the lattice. A static antiquark is introduced by the complex conjugate variable.

The change in the free energy of the ensemble caused by the addition of a single heavy quark is given by

$$\exp[-f(T,m_q)/T] = \langle P(0)/3 \rangle_U \qquad (30)$$

### 2.4.4. Heavy Quark Potential

The thermal heavy quark potential is determined from

$$\exp[-V(r,T)/T] = \langle P(0)P^\dagger(r)/9 \rangle_U \qquad (31)$$

The quantity $V(r,T)$ is more precisely the change in the free energy of the ensemble caused by adding a spinless quark and antiquark pair at separation $r$.

## 3. Phase Structure

In nature the quark masses assume their physical values, of course. In numerical simulations, however, it is possible to adjust quark masses and other parameters to gain more insight into the phase structure of QCD. Thus a phase diagram can be constructed in the multidimensional space of the temperature $T$, the quark masses $m_i$ for $i = 1, \ldots, N_f$ flavors and the corresponding chemical potentials $\mu_i$. The majority of simulations have been carried out at zero chemical potential with two or four flavors of equal mass quarks. However, there are a few simulations that approach a more physical quark mass spectrum with two equal mass light quark flavors and one strange quark. Because simulations become more costly as the quark mass is decreased, simulations are done at an unphysically large quark mass, requiring an extrapolation to smaller physical values.

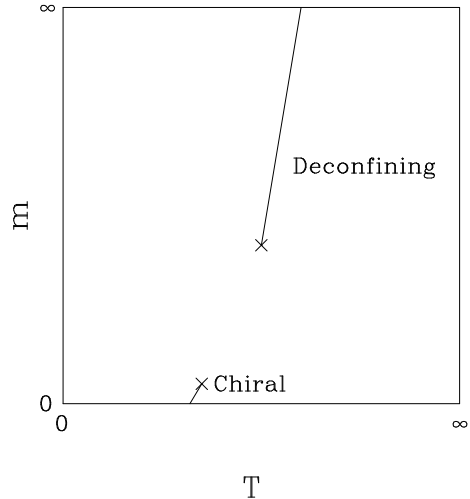

Figure 1: Schematic phase diagram in quark mass and temperature for QCD.

At infinite mass, quarks are absent from the thermal ensemble, and QCD becomes a pure SU(3) Yang-Mills theory. It is well established that this theory has a first order nonzero temperature phase transition. Indeed the attendant coexistence of plasma and confined phases at the critical temperature inspired speculation about phase boundaries and bubbles in plasma production and cooling. At the opposite extreme of zero quark mass, QCD becomes a chirally symmetric theory. The symmetry is spontaneously broken at low temperature, leading to a zero mass pi meson. As temperature is increased, a phase transition occurs, leading to restoration of the broken symmetry and the appearance of chiral multiplets in the hadron spectrum. Thus we have the schematic phase diagram for QCD as a function of a single $SU(N_f)$ flavor symmetric quark mass and temperature shown in Fig. 1. The first order confinement/deconfinement phase boundary has been extended to finite quark mass. Whether the chiral phase boundary extends to nonzero quark mass as shown or occurs only at zero mass depends on the number of flavors. With three or four light quarks quite general arguments can be made in favor of a first order chiral phase transition.[14] In that case an extended chiral phase boundary is expected. But with two light flavors a second order phase transition is not ruled out,[14] in which case it may occur only at zero quark mass. Thus symmetry and universality considerations do not require a phase transition at physical values of the quark masses. To establish its existence, we turn to numerical simulation.

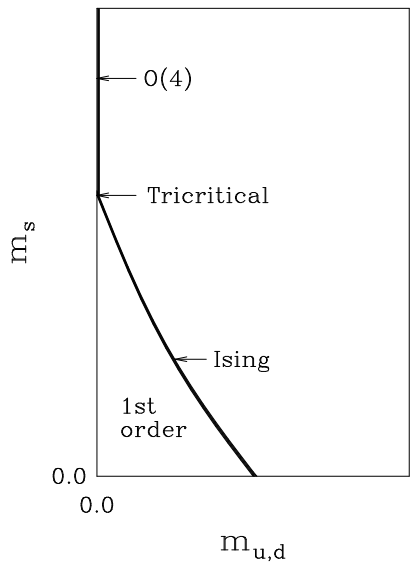

Figure 2: QCD phase diagram for $2+1$ flavors as a function of a degenerate up and down quark mass $m_{u,d} = m_u = m_d$ and a strange quark mass $m_s$. The heavy line is the critical phase boundary.

Are these expectations confirmed in numerical simulations? For four flavors of equal mass quarks numerical evidence suggests that the chiral phase boundary extends to nonzero quark mass, but it is likely that a gap separates the chiral and deconfining phase boundaries as sketched in Fig. 1.[3,15] For two flavors of equal mass quarks there appears to be no phase transition at a small, but nonzero quark mass.[16,24] For two equal light quark masses and one strange quark, we view the phase diagram from a different perspective in Fig. 2.[17] Such a phase diagram is motivated by an analysis of the corresponding sigma models in mean field theory, augmented by an analysis of quantum fluctuations.[14,18,19] The diagram indicates for which range of the two mass parameters a nonzero temperature chiral phase transition occurs and whether the phase transition is expected to be first or second order. When the strange quark mass is large, the thermal ensemble is effectively a two-flavor ensemble, and we have a second order phase transition only at zero up and down quark mass. However, when the strange quark is sufficiently light, we recover the first order chiral phase transition expected of the three-flavor ensemble, which extends to nonzero quark mass as sketched. Whether QCD conforms to this expectation in detail remains to be established.

So is there a phase transition at physical values of the quark masses? Simulations with $2+1$ flavors in the staggered and Wilson fermion schemes both support the existence of the first order region sketched in Fig. 2, but do not agree on its extent.[20,21,22] Staggered fermion simulations of the Columbia group found a first order signal for $(m_{u,d}, m_s) \approx (15, 15)$ MeV, but none for $(m_{u,d}, m_s) \approx (15, 30)$ MeV, suggesting no phase transition at physical values. A recent Wilson fermion simulation by the Tsukuba group found a first order signal with quark masses as large as $(m_{u,d}, m_s) \approx (140, 140)$ MeV and $(m_{u,d}, m_s) \approx (0, 400)$ MeV,[20,22] allowing a phase transition at physical values. Although the staggered fermion approach for studying nonzero temperature QCD is more mature than the Wilson approach, until algorithmic improvements lead to consistency, we cannot be certain whether there is a phase transition. Nonetheless, it is likely that if there is a phase transition at all, it is weak. Still, as we shall see, even if there is no phase transition, there is a dramatic crossover.

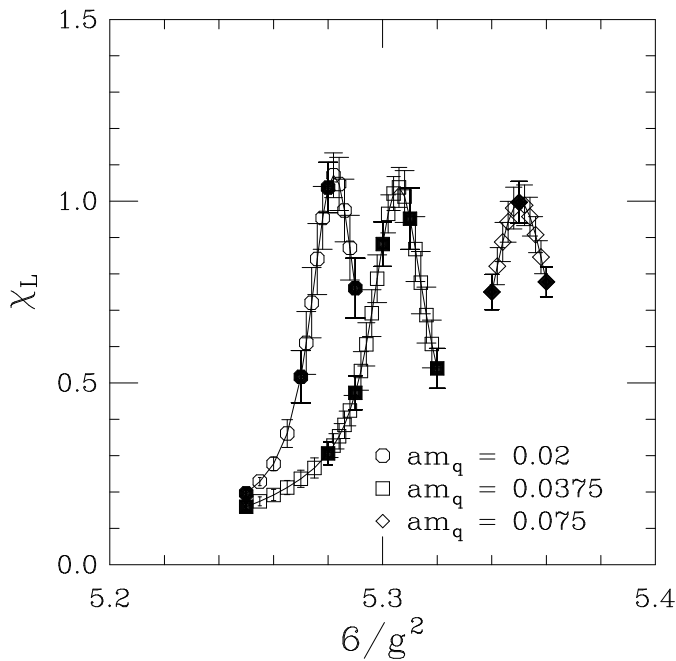

Figure 3: Polyakov loop susceptibility *vs.* $6/g^2$ from Karsch and Laermann. Solid symbols are directly simulated. Open symbols are derived by reweighting. Peaks locate the crossover.

## 4. Temperature of the Phase Transition

As we have seen, light quarks make a significant difference in the character of the phase transition, so cannot be neglected in a realistic study of thermodynamics. The temperature of the phase transition or crossover in the presence of dynamical quarks is also an important indicator of convergence to the continuum limit. Knowing its approximate value as well is of considerable phenomenological importance.

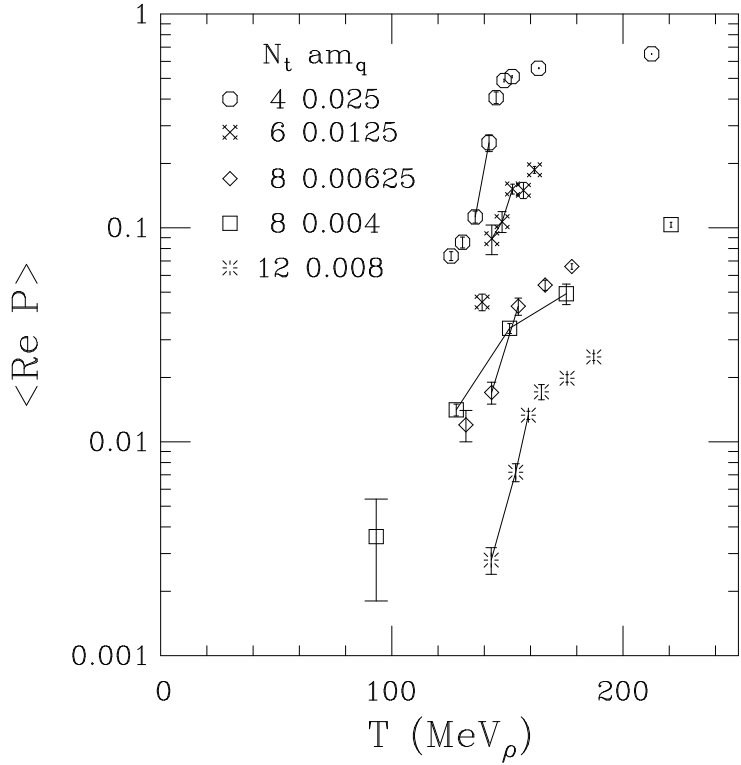

Figure 4: Polyakov loop *vs.* temperature from Ref. 27. The line segments indicate the range of uncertainty in locating the maximum slope. Data are from Refs. 24,26,30,37,26.

There are two methods to locate the crossover. The most elegant locates the peak in a susceptibility, such as that based on the Polyakov loop.[30]

$$\chi_L = N_s^3[\langle(\mathrm{Re}P)^2\rangle - \langle\mathrm{Re}P\rangle^2]. \tag{32}$$

Combined with a Ferrenberg-Swendsen reweighting analysis to interpolate between simulation points, the method permits a clean determination of the crossover coupling, as shown in Fig. 3. A second method locates the peak in the derivative of a sensitive observable, as illustrated for the Polyakov loop variable in Fig. 4.

The equilibrium temperature in a lattice simulation is the reciprocal of the lattice extent in the imaginary time dimension, or $1/N_t a$. To express the crossover temperature in physical units requires the measurement of an additional quantity at the same lattice parameters where the crossover is observed, for example, the (zero temperature) rho meson mass, also determined in dimensionless lattice units as $m_\rho a$.

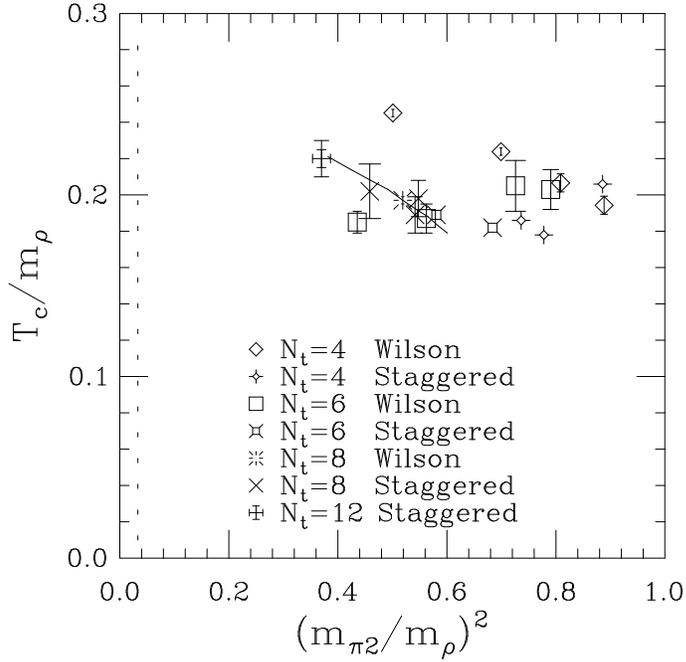

Figure 5: From Ref. 24. Temperature of the crossover in units of the rho meson mass with two light quarks *vs.* the squared ratio of the pion to rho mass. For points from the staggered simulations, the mass of the second (non-Goldstone) pion is used. The curved line segment shows the error bar for the new $N_t = 8$ staggered point. The vertical dashed line indicates the physical mass ratio. Data are from Refs. 24,26,37,50

Thus
$$T_c/m_\rho = 1/(N_t a m_\rho) \tag{33}$$
To set the scale in this way requires a separate simulation at low temperature. From the experimental value the temperature is then converted to physical units:
$$T_c(\text{MeV}_\rho) = 770 T_c/m_\rho = 770/(N_t a m_\rho) \tag{34}$$
The result must be interpreted cautiously, however, since in present simulations with dynamical quarks, the ratio of the nucleon to rho meson mass is unphysically high. Setting the scale with the nucleon mass instead would therefore result in a still lower temperature. There is some hope for future improvement: Through a series of extrapolations it has been possible to reach physical values of the nucleon to rho mass ratio in recent zero temperature quenched calculations for hadronic states constructed from both Wilson and staggered valence fermions.[23] The same accomplishment with dynamical quarks included will take considerably more effort. Another problem affecting the determination of the crossover temperature is the need for a rather long extrapolation to physical quark masses. In present simulations with dynamical quarks the zero temperature mass ratio $m_\pi/m_\rho$ lies in the range 1/3 to 1/2, twice to three times the experimental value of 0.18. This ratio can be reduced to its physical value by reducing the quark mass considerably. So for now we use the notation MeV$_\rho$ to call attention to the assumptions made.[26]

The temperatures thus determined are plotted as a function of the square of the ratio of the pi to rho mass in Fig. 5.[24] This figure includes results from simulations in the Wilson as well as staggered fermion schemes. As we have noted, the Wilson scheme breaks chiral symmetry explicitly and the staggered fermion scheme breaks the continuous flavor symmetry explicitly. The breaking of these symmetries results in an unphysical value for the chosen ratio. Both symmetries are expected to be restored in the continuum limit, permitting approach to the physical value. Thus progress toward the continuum limit is measured by movement toward the physical ratio. There is an encouraging consistency in the results, which place the crossover temperature at approximately

$$T_c = 140 - 160 \text{ MeV}_\rho. \tag{35}$$

By contrast, in a gluon ensemble without quarks (quenched approximation), the temperature of the phase transition is about 260 MeV.[25] For further details concerning the construction of the temperature scale, please see Refs. 26,27.

## 5. Consequences of Chiral Symmetry Restoration

The QCD phase transition or crossover inherits characteristics of both the deconfinement and chiral phase transitions. An immediate consequence of chiral symmetry restoration is the vanishing of the chiral order parameter $\langle \bar\psi \psi \rangle$ in the limit of zero quark mass. This effect is observed. Moreover, the vanishing is quite abrupt. We begin this section with a discussion of these results. The appearance of critical behavior is a second consequence of chiral symmetry restoration, depending sensitively on the values of the up, down, and strange quark masses.[19] Critical behavior is expected along the curve shown in Fig. 2. However, since so little work has yet been done with two plus one flavors, we restrict our discussion of critical behavior to the case of zero quark mass in two-flavor simulations. Finally, the restoration of chiral symmetry also produces dramatic shifts in the hadron spectrum, leading to the emergence of chiral multiplets. These multiplets appear in long range screening correlators, e.g. the high temperature Yukawa interaction between two nucleons and are readily observed in numerical simulations. They do not appear to have any directly observable consequences, but their existence imposes constraints on phenomenological models.

### 5.1. Chiral Order Parameter

The dramatic falloff in the chiral order parameter $\langle\bar\psi\psi\rangle$ has long been used to locate the high temperature crossover. Figure 6 from Boyd et al.[28] gives a compilation of results for the ratio $\langle\bar\psi\psi\rangle(T)/\langle\bar\psi\psi\rangle(T=0)$. This figure collects results for a variety of staggered quark flavors, including quenched ($N_f = 0$) at $N_t = 4$[29] and $N_t = 8$,[28] two flavors $am_q = 0.02$,[30] three flavors (two light, one heavier) with $am_{u,d} = 0.0125$ and $am_s = 0.25$,[31] and four flavors.[32] The quenched results are extrapolated to zero quark mass. The others are at small quark mass, but not extrapolated to zero. The zero temperature value is from Ref. 33.

Also evident in Fig. 6 is a remarkable constancy in the order parameter for $T < 0.92T_c$. The same study also found no significant variation in the pion decay constant $f_\pi$ over this temperature range.[28] Thus any experimentally observed shifts in hadron properties would indicate temperatures very close to or above the critical temperature.

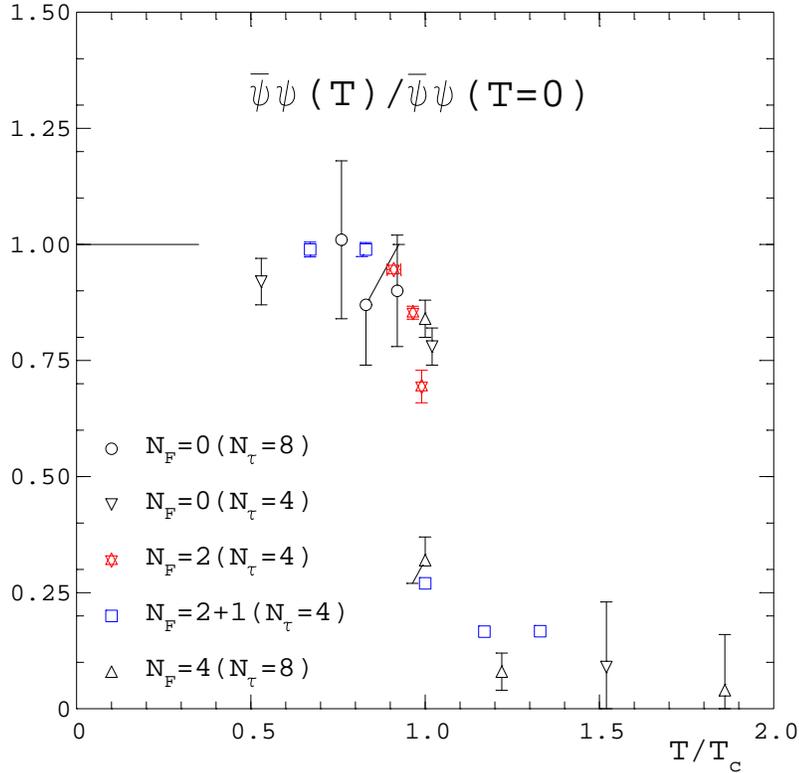

Figure 6: From Ref. 28. Chiral condensate normalized to its zero temperature value as a function of $T/T_c$. Zero flavor results are extrapolated to zero quark mass. The others are not. See text for details.

## 5.2. Critical Behavior

If QCD falls into the proposed universality class of sigma models with a second order chiral phase transition at the boundaries indicated in Fig. 2, certain scaling relations apply near the critical point.[34] The question is crucial for a successful extrapolation to small quark mass and large volume near the phase transition. A correct extrapolation requires the correct critical exponents. Thus an important test of the proposed phase structure is to determine whether QCD has the expected critical behavior. To emphasize that the conclusion is not foregone, a recent analysis of the three-dimensional Gross-Neveu model by Kocić and Kogut questions the conventional wisdom that places QCD with its composite scalar mesons in the same universality class as sigma models with their elementary scalar mesons.[35] Instead, in a detailed simulation Kocić and Kogut found mean-field scaling.

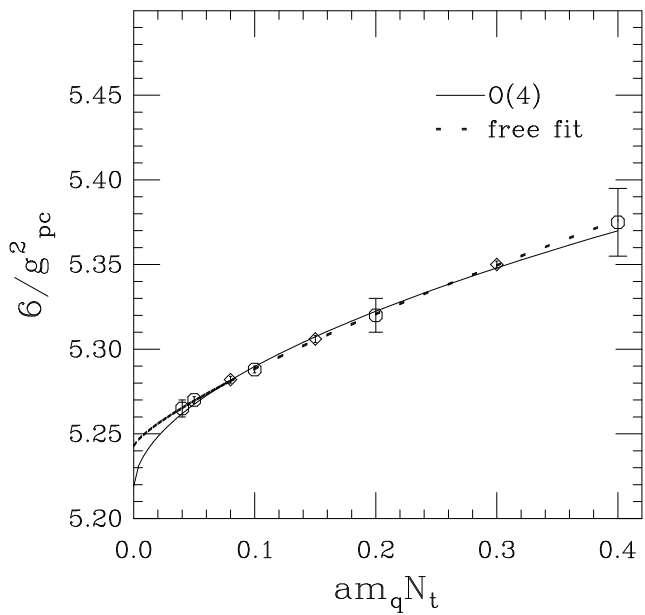

Figure 7: Crossover coupling *vs.* quark mass for two flavors of staggered quarks from Ref. 30. Solid line is a fit with O(4) critical exponents. Dashed line is a free fit.

Let us focus on the two flavor theory in the staggered fermion scheme, corresponding to the upper portion of the $m_s$ axis in Fig. 2. In this case O(4) universality is expected in the continuum limit. To be more precise, in the staggered fermion scheme, one expects O(2) critical behavior on coarse lattices where the flavor sym-metry breaking of the staggered scheme is significant, and O(4) as the continuum limit is approached. Recent work by Karsch and Karsch and Laermann attempts a determination of some of the critical exponents of QCD.[30,36] They exploit the standard sigma model analogy between QCD and a magnetic system. In this analogy the quark mass plays the role of an external magnetic field and $\langle \bar{\psi}\psi \rangle$ plays the role of the magnetization.

The standard analysis of critical behavior begins with the assumed scaling of the critical contribution to the free energy in a magnetic system.[34] This contribution is singular and dominant at small field close to the zero field critical temperature $T_c(0)$. The scaling property, expressed in terms of the scaled temperature $t$,

$$t = [T - T_c(0)]/T_c(0) \tag{36}$$

and magnetic field $h$ is

$$f_{\text{crit}}(t, h) = b^{-1} f_{\text{crit}}(b^{y_t} t, b^{y_h} h). \tag{37}$$

From this scaling behavior one derives a scaling relation for the critical contribution to the magnetization $s = -\partial f_{\text{crit}}/\partial h$:

$$s(t, h) = h^{1/\delta} y(x) \tag{38}$$

where

$$x = t h^{1/\beta\delta} \tag{39}$$

and $y(x)$ is a scaling function. Here $\beta$ and $\delta$ are the critical exponents appropriate to the universality class. An important consequence of this result is that a crossover peak in the susceptibility $\chi_h = \partial s/\partial h$ moves along a curve of constant $x = x_{\text{pc}}$ as $h$ and $t$ are varied. Thus if critical scaling holds, once the pseudocritical temperature is known at one $h$, it can be predicted at all $h$.

In QCD the quark mass plays the role of the magnetic field and $\langle \bar{\psi}\psi \rangle$, the magnetization. Specifically, Karsch suggests the identification

$$h = m_q/T = a m_q N_t \tag{40}$$
$$t = 6/g^2 - 6/g_c^2(0, N_t), \tag{41}$$

where $g_c(0, N_t)$ is the critical gauge coupling at zero quark mass for a particular $N_t$.[36] Scaling then predicts how the critical gauge coupling changes as a function of the quark mass:

$$6/g_{\text{pc}}^2(m_q a) = 6/g_c^2(0) + (m_q/T)^{1/\beta\delta}. \tag{42}$$

Using this expression Karsch presented an analysis of the crossover for $N_t = 4$, 6, and 8 for two quark flavors.[36] The agreement is quite encouraging. The addition of new data at $N_t = 4$ permits a more refined analysis, shown in Figure 7.[30] Their best fit critical exponent $1/\beta\delta$ is $0.77 \pm 0.14$, in slight disagreement with the O(4) value 0.55(2), but consistent with the O(2) value 0.60(1) and the mean field value 0.67.

Karsch and Laermann also introduced a new cumulant

$$\Delta = \frac{\partial \ln\langle\bar\psi\psi(6/g^2,m_q)\rangle}{\partial \ln m_q} = \frac{1}{\delta} - \frac{xy'(x)}{\beta\delta y(x)} \tag{43}$$

that evaluates the critical exponent $\Delta = 1/\delta$ at $x = t = 0$. They obtain $0.21 < 1/\delta < 0.26$ consistent with the O(4) value 0.208(2) and O(2) value 0.2080(3), and somewhat inconsistent with the mean field value 0.33. Thus the Kocić-Kogut scenario cannot be decisively excluded.

To push these results further we can test the full scaling relation (38) in simulations with two flavors of staggered fermions over the wide range of currently available $N_t$. For this purpose we use slightly different variables to permit comparison among different $N_t$ and to avoid quantities with anomalous dimensions, namely,

$$h = m_\pi^2(m_q, T=0)/m_\rho^2(m_q, T=0) \tag{44}$$
$$t = [T - T_c(0)]/T_c(0) \tag{45}$$
$$s = h^{-1} m\langle\bar\psi\psi(m_q, T)\rangle/T^4 \tag{46}$$

The scaling relation (38) then gives a universal function

$$y(x) = h^{-1-1/\delta} m_q \langle\bar\psi\psi(m,T)\rangle/T^4 \tag{47}$$

with $x = t h^{1/\beta\delta}$. The extra factor $h^{-1}$ is needed to compensate for the quark mass factor $m_q$.

To implement the alternate variables (46) one must know the zero temperature pi and rho masses over the parameter range of the nonzero temperature analysis. This is done by constructing an interpolation of known hadron masses.[26,27]

This analysis was applied to data for $\langle\bar\psi\psi\rangle$ from several groups.[37] Setting the critical exponents $\delta$ and $\beta$ to their O(4) values and adjusting the sole remaining free parameter $T_c(0)$ in 10 MeV increments to get the best agreement gives the result shown in Fig. 8 for $T_c(0) = 150$ MeV. With the exception of the $N_t = 12$ data, the scaling agreement is rather good. At this level it is not possible to distinguish O(4) from O(2) and mean field critical behavior. Setting $T_c(0)$ to 140 MeV or 170 MeV worsens the agreement noticeably, but 160 MeV gives comparable consistency. Obviously a host of systematic errors, including finite volume effects and deviations from continuum scaling enter the analysis, so refinements are certainly needed before the method can serve as a definitive test of critical behavior. The most glaring inconsistency in this figure comes from the preliminary $N_t = 12$ data. Increasing $T_c(0)$ for only this data set to 160 MeV and plotting it with other data computed for $T_c(0) = 150$ MeV brings the $N_t = 12$ data at the lower quark mass into good agreement. Thus the discrepancy could be caused either by a gradual upward shift in the crossover temperature as the lattice spacing is decreased, or by an erroneous extrapolation of the hadron spectrum above $6/g^2 = 5.7$, or by a failure of the scaling hypothesis over this parameter range.

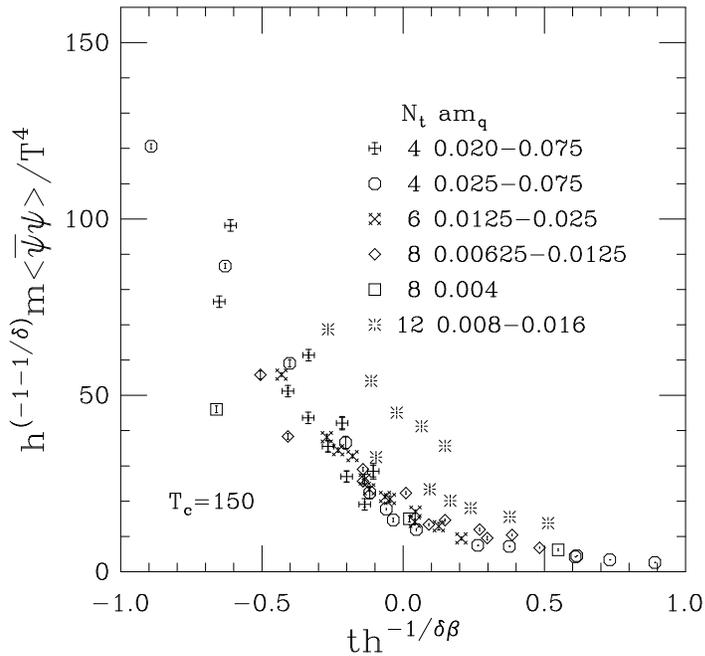

Figure 8: Scaled $\langle\bar{\psi}\psi\rangle$ vs. scaled temperature with O(4) critical exponents. Data are from Refs. 24,26,30,37.

5.3. Screening Spectrum

When a symmetry is restored in a hamiltonian system, we expect eigenstates to organize themselves into multiplets belonging to representations of the symmetry group. The chiral group of QCD is $U_A(1) \times SU(N_f) \times SU(N_f)$, depending on the number of degenerate quark flavors $N_f$. At low temperature the flavor singlet $U_A(1)$ symmetry is explicitly broken by the gauge anomaly and the flavor nonsinglet symmetry $SU(N_f) \times SU(N_f)$ is spontaneously broken to the flavor group $SU(N_f)$. The gauge anomaly receives contributions from instanton gauge configurations, which are expected to be suppressed at high temperature.[38] The spontaneously broken nonsinglet symmetry is also expected to be restored at high temperature. If chiral symmetry is restored, which part of the symmetry is restored?[39,40] The answer to this question helps us understand what soft modes may be present at the crossover. Critical fluctuations drive soft modes, and may give rise to collective behavior with observable consequences.[41] We may seek the answer by examining the spectral multiplets. For example, under $SU(2) \times SU(2)$ the expected low-lying meson multiplets are the quartets $\{\pi, \sigma\}$ and $\{\eta, a_0\}$ and the sextet $\{\rho, a_1\}$ among others. On the other hand under the $U_A(1)$ symmetry all flavor multiplets are parity doubled, so the low-lying multiplets are the sextets $\{\pi, a_0\}$ and $\{\rho, a_1\}$ and the doublet $\{\sigma, \eta\}$ among others.

Ordinarily we would examine the multiplet degeneracy by measuring hadron masses. These are obtained by measuring hadron propagators between points widely separated in Euclidean time. At nonzero temperature, however, the Euclidean time dimension is limited, so we measure hadron propagation between points widely separated in the spatial direction instead. Such propagators are often called "screening propagators" or "screening correlators". For example, from the generic meson propagator

$$G_{ab}(r) = \langle H_a(0) H_b(r) \rangle, \qquad (48)$$

where $H_a(r) = \bar{\psi}(r) \Gamma_a \psi(r)$ with $r = (x, y, z, \tau)$, we may construct the static correlator by integrating over all but the $z$ coordinate:

$$C_{ab}(z) = \delta_{ab} C_a(z) = \delta_{ab} \int d\tau dx dy G_a(x, y, z, \tau) \qquad (49)$$

This correlator receives contributions from all possible states in a given channel. Thus the asymptotic behavior is given by

$$C_a(z) = \sum_{n=0}^{\infty} \rho_{an} \exp(-m_{an} z) \qquad (50)$$

At large enough $z$ the correlator falls off as

$$C_a(z) \sim \rho_{a0} e^{-m_{a0} z} \qquad (51)$$

where $m_{a0}$ is the lightest screening mass for the particular channel. These are the masses that may be used for exploring the multiplet structure.

The screening propagators have a physical interpretation. For example, the pion screening state is the exchanged object that give rise to the Yukawa interaction between two static nucleons in the thermal ensemble.

The screening masses have been popular observables in nonzero temperature simulations for some time.[39,42,43,44] An example showing the spectrum as a function of temperature is given in Fig. 9. The state labeled $\sigma$ in the figure is really only the valence quark component, corresponding to the first term in Eq. (28). Let us call it $\sigma_v$.[39]

It is apparent that rapid changes take place in the screening spectrum at the phase transition. Can we tell from the multiplet structure which chiral symmetry is restored? To answer the question, we must take the chiral limit, i.e., take the quark masses to zero. The result appears to be consistent with the formation of an SU(2)×SU(2) $\{\pi, \sigma_v\}$ multiplet among others, suggesting restoration of SU(2)×SU(2). In these simulations the other $\{\eta, a_0\}$ multiplet was not studied, so we can't tell whether the $U_A(1)$ symmetry is restored simultaneously. But a complication arises. The valence part $\sigma_v$ is not distinguishable from an $a_0$ in the continuum limit.[40] That would make the observed multiplet structure $\{\pi, a_0\}$, consistent with $U_A(1)$ restoration. However, it is argued in Ref. 39 that in the chiral limit a chiral selection rule eliminates the second (nonvalence) contribution to the $\sigma$ propagator, leaving us a multiplet structure consistent with the restoration of either symmetry. Further work is evidently needed to resolve the question.

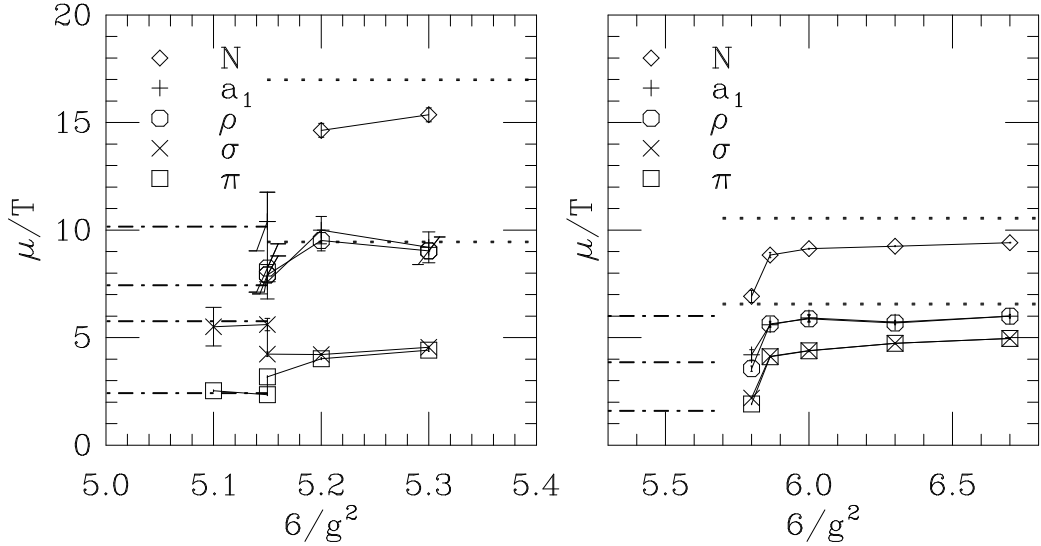

Figure 9: Screening masses as a function of $6/g^2$ (hence temperature).[44] Results for four flavors of staggered quarks on a $16^3 \times 8$ lattice with $ma = 0.01$ are shown in the left panel. The phase transition takes place at $6/g^2 = 5.15$. The dot-dashed lines on the left mark from top to bottom, respectively, the zero temperature masses in units of $T_c$ for the nucleon and $\rho$, $\sigma$, and $\pi$ mesons. The dotted lines on the right indicate from top to bottom the screening masses expected from propagation of three and two free quarks. Results for a zero flavor (quenched) simulation on a $16^3 \times 4$ lattice are shown similarly in the right panel. The points at $6/g^2 = 5.865$ are from Ref. 43 and correspond to $T = 1.5T_c$. The dot-dashed lines mark the zero temperature masses of the nucleon, $\rho$, and $\pi$. The free quark values differ in the two panels because of the different lattice sizes.

Shown on the high temperature side of the phase transition in Fig. 9 is the result expected from a continuum consisting of two and three free quarks. It is remarkable that most of the screening masses are in rough agreement with these values. In particular, the $\rho$ and $a_1$ screening masses are close to the two-quark continuum and the nucleon mass is close to the three-quark continuum. It is clear from Fig. 9 that only the pion deviates significantly from the free-quark rule. Although such a result suggests deconfinement in the high temperature phase, the spectrum alone isn't decisive: charmonium has a mass close to twice the charm quark mass, but it is confined.

The continuum value of the free quark screening mass is simply

$$m_{\rm eff} = \sqrt{(\pi T)^2 + m_q^2}, \tag{52}$$

where the Lagrangian quark mass is $m_q$. This result follows from the requirement that the quark field be antiperiodic in Euclidean time. Therefore the minimum time component of the quark and antiquark momentum is $\pi T$ and it propagates in the spatial direction with the stated effective mass. For sufficiently light quarks, the quark-antiquark threshold in the meson screening channel is $2\pi T$. For the baryon channel the three-quark threshold is $3\pi T$. These continuum values are modified on a finite lattice, as shown in the figure.[44]

Are these high temperature screening states merely a multiquark continuum? We return to this question in Sec. 9.

## 6. Thermodynamics with Wilson Fermions

To be confident that numerical simulations accurately represent continuum QCD it is essential that we demonstrate that our answers are independent of the fermion scheme. Unfortunately, thermodynamic simulations with Wilson fermions are not sufficiently developed at present to make a credible comparison with the staggered scheme. The fundamental difficulty is that we are dealing with a phase transition associated with the restoration of a spontaneously broken chiral symmetry, but the Wilson scheme builds in an explicit breaking of this symmetry, which goes away only in the continuum limit.

For a given lattice dimension the staggered fermion parameter space consists of the gauge coupling $6/g^2$ and the quark masses $ma$. The chiral limit is reached at zero quark mass. In the Wilson fermion scheme each quark mass is replaced by a hopping parameter $\kappa$. The chiral limit is not known a priori, but in simulations at low temperature and reasonably high values of $6/g^2$, it is found that a number of indicators of chiral symmetry, e.g. a vanishing zero temperature pion mass and a vanishing current quark mass, hold at least with some consistency along a "chiral curve" in the parameter space

$$\kappa = \kappa_c(6/g^2). \tag{53}$$

At strong coupling (small $6/g^2$) there is no assurance that there is any consistency among the indicators of chiral symmetry, so they must be checked in simulations.

The thermal crossover occurs along another line depending on $N_t$

$$\kappa = \kappa_t(6/g^2, N_t). \tag{54}$$

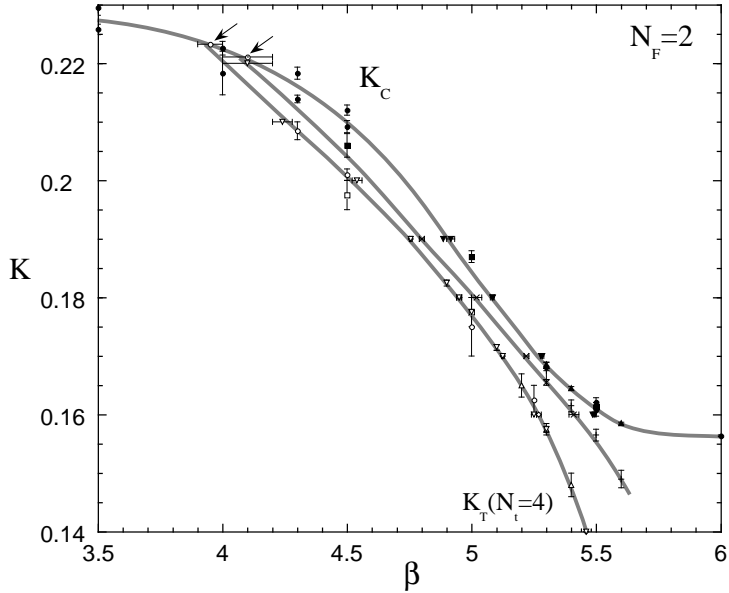

Figure 10: Phase diagram for two flavors of Wilson fermions from Ref. 20 as a function of the Wilson hopping parameter $\kappa$ and coupling $\beta = 6/g^2$. The top curve $\kappa_c$ locates the chiral line. The bottom curve gives the thermal crossover $\kappa_t$ for $N_t = 4$ and the middle curve gives the thermal crossover $\kappa_t$ for $N_t = 6$. In each case the low temperature phase is below the $\kappa_t$ curve and the high temperature phase is above. The arrows indicate the intersection of the thermal and chiral curves in each case as reported by the Tsukuba group. Data are from Refs. 49,46,47,48,45.

The chiral line and thermal lines for $N_t = 4$ and $N_t = 6$ are indicated in Fig. 10. As the chiral line is approached from the low $\kappa$ side, the quark mass is decreasing toward its physical value. We are interested in studying the crossover or phase transition in a region of small (physical) quark mass, so close to the chiral line. To avoid lattice artifacts, we would also like to work at a small lattice spacing. Indeed, evidence for lattice artifacts in two flavor simulations at $N_t = 6$ was recently reported by the MILC collaboration, which found a first order, possibly bulk, phase transition in simulations at $\kappa = 0.17, 0.18$, and $0.19$ and $6/g^2$ in the range $4.8 - 5.2$. in close proximity to the thermal crossover.[47] Thus one might infer that it would be preferable to work with values of $6/g^2$ larger than about 5.

An unfortunate feature of the phase diagram, evident in Fig. 10, is that with $N_t = 4$, before we reach the chiral limit at large $6/g^2$, we must first cross to the high temperature phase.[48] In other words, to study the crossover at small quark mass, we must work near the intersection of the thermal and chiral lines. The Tsukuba group discovered that for $N_t = 4$ the intersection occurs at very small values of $6/g^2$, usually considered to be in the strong coupling regime. At $N_t = 6$ the intersection occurs at slightly larger $6/g^2$, as we can see in Fig. 10.[49]

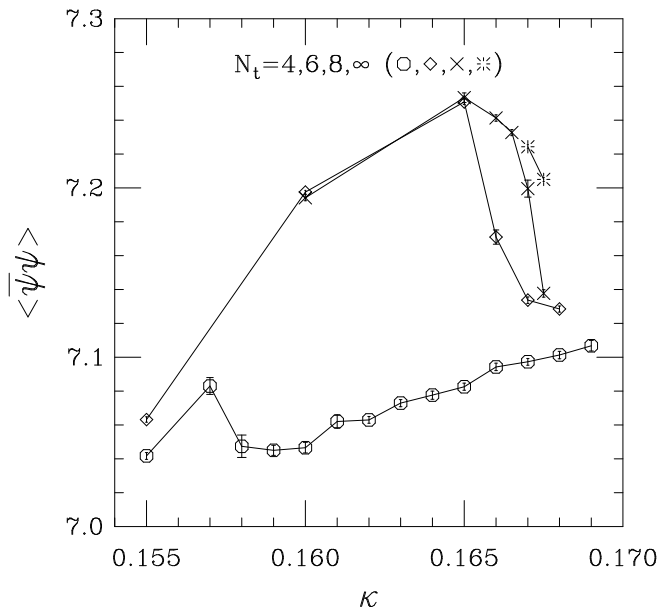

Figure 11: From Ref. 50. Chiral condensate *vs.* $\kappa$ and various $N_t$ for two flavors of Wilson fermions at $6/g^2 = 5.3$. Data are from Refs. 45,47,50.

At $N_t = 8$, not shown in the figure, another shift in the right direction is found.[50] A simulation at $N_t = 8$ was done at $6/g^2 = 5.3$ over a range of $\kappa$ up to $\kappa_c \approx 0.168$, allowing a comparison with previous results at $N_t = 4$ and 6. The thermal crossover, now shifted to $\kappa_t \approx 0.167$, shows no evidence for the presumed lattice artifact seen at slightly larger $\kappa$ at $N_t = 6$. As illustrated for $\langle \bar\psi \psi \rangle$ in Fig. 11, as $N_t$ is increased, bulk quantities appear to follow an envelope established by the zero temperature theory, breaking away at the crossover. Moreover $\langle \bar\psi \psi \rangle$ appears to be decreasing immediately prior to the crossover, suggesting progress toward a low quark mass.[50]

So, although the thermal line continues to shift in the right direction at $N_t = 8$, it is still not enough. Apparently it is necessary to work at quite high $N_t$, if we want a simulation that is as close to the continuum limit as the staggered fermion simulations whose predictions we are trying to confirm.[49] Thus with the original Wilson action, we are forced to chose between a strong coupling simulation and risk encountering lattice artifacts or carry out an expensive simulation. Further progress with the Wilson scheme is likely to require working with an improved action. The Tsukuba

group has adopted one such improvement with encouraging preliminary results.[20,51]

## 7. Structural Changes at the Crossover

Dramatic changes take place at the thermal crossover. Hadrons grow and merge, resulting in an extended mixture of quarks and gluons. Understanding the crossover is vitally important to the development of models of hadronization. In this section we examine two indicators of structural change in the QCD ensemble: the "constituent quark free energy" and the baryon density induced by the introduction of a point color triplet. We also consider the heavy-quark potential in a pure gluon ensemble. Finally we mention briefly recent efforts to explain the crossover in terms of topological structures, namely instantions and monopoles.

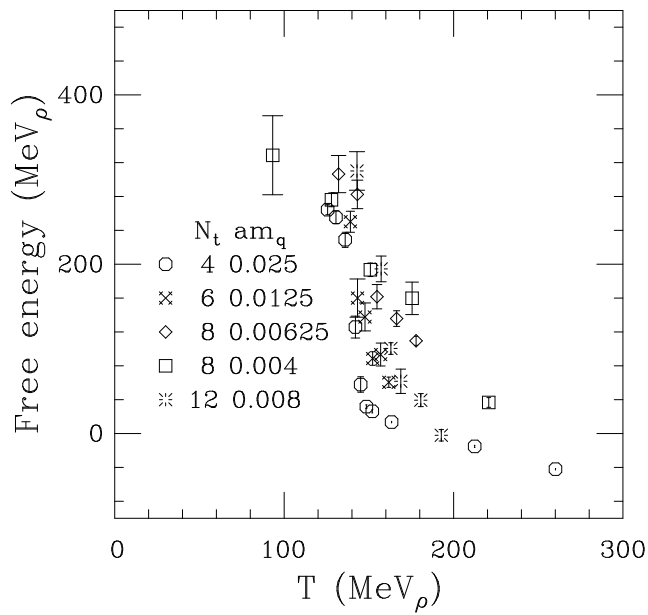

Figure 12: Constituent quark free energy as defined in text with $c$ adjusted to 0.5. Data are from the same sources as Fig. 4.

### 7.1. Constituent Quark Free Energy

When a point color triplet (a spinless test quark) is placed in the thermal ensemble at low temperature, we expect that a gluon cloud surrounds it and a light antiquark binds to it to form a color singlet. This effect is induced in the ensemble by the Polyakov loop operator. The expectation value of the Polyakov loop variable measures the change in the free energy of the thermal ensemble due to the introduction of a point spinless test quark. This free energy difference, Eq. (30)

$$f(T, m_q) = -T \log \langle \text{Re} P/3 \rangle_U, \qquad (55)$$

is a function of the temperature $T$ and light quark mass $m_q$. At low temperature, the point source is screened by a light antiquark, forming a sort of heavy-light meson. The free energy $f(T, m_q)$ then consists of the energy of the screening cloud and the lattice-regulated ultraviolet-divergent self-energy of the point source. The former contribution might be called loosely the "constituent quark free energy". At high temperature, screening is accomplished through thermal fluctuations in the color fields. The point source self-energy still diverges in the same way as at low temperature, however. Thus if we could isolate the contribution to $f(T, m_q)$ from the screening cloud, we would obtain the constituent quark free energy as a function of temperature, which may be of some interest in constructing models of the quark plasma.

To compute the self energy of an isolated point source requires introducing an infrared cutoff, which can be temperature dependent. This requirement introduces an element of arbitrariness in the definition. In simplest terms, we must decide where to put the knife when we dissect the heavy meson.

For an initial stab at this analysis, we observe that the regulated ultraviolet divergence is proportional to $1/a$. Thus we have

$$f(T, m_q) = f_{\text{cq}}(T, m_q) + c/a \qquad (56)$$

for some constant $c$. Since $N_t = 1/aT$, we solve for the constituent quark free energy

$$f_{\text{cq}}(T, m_q) = -T \left( \log \langle \text{Re} P/3 \rangle + c N_t \right). \qquad (57)$$

This expression determines the scaling of the Polyakov loop measurement as $N_t$ is varied. If the dimensionless constant $c$ is chosen properly, and if continuum scaling holds, Polyakov loop data at increasing $N_t$ should yield a universal function $f_{\text{cq}}(T, m_q)$.

The treated Polyakov loop data for a wide range of $N_t$ is plotted in Fig. 12. The constant $c$ is adjusted by eye to achieve the rough scaling agreement shown. For this purpose only values for the lightest available quark mass from each data set are used. Although the quark mass values $m_q/T$ are not the same from one $N_t$ to the next in this figure, they are small ($m_q/T \leq 0.1$) and would be expected to contribute little (of the order 10 MeV) to the free energy. Thus one would expect only a small inconsistency from this variation. It is amusing that at the crossover, the free energy drops by about the 300 MeV expected in a constituent quark model with deconfinement at high temperature.

7.2. Induced Baryon Number

Useful insights into the structure of the plasma can be obtained by measuring the quark number density in the vicinity of the point test charge. From our discussion of the constituent quark self energy, we observed that at low temperature the screening cloud very likely contains an antiquark. Two light quarks may also combine with the point charge to form a baryon-like color singlet, but (at least we would expect) with lower probability, since the Boltzmann factor should suppress this contribution. Crudely, the suppression should go as $\exp(-m_{\rm const}/T)$ where $m_{\rm const}$ is the mass of the extra constituent quark. At crossover, this mass is nearly twice the temperature, which would give an order of magnitude suppression. However, things are changing rapidly at the crossover. We learn more about the effect by measuring the dynamical quark number density $\rho_q(r)$ in the presence of the test charge as a function of the distance from the test charge.[52] The total induced quark number

$$Q = \int d^3 r \rho_q(r) \qquad (58)$$

gives another measure of the screening cloud.

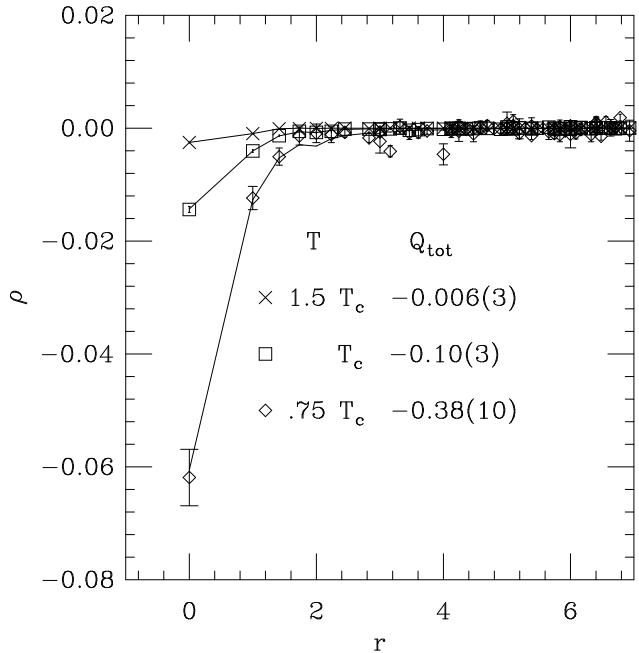

Figure 13: Quark number density induced by a fixed quark at the origin as a function of distance from the origin at three temperatures. Curves are fits to a single screening mass. The total induced quark number $Q$ is also shown.

Results of simulations with two light quark flavors at three temperatures, $0.75T_c$, $T_c$, and $1.5T_c$ are shown in Fig. 13.[52]. We see that as the temperature is increased, the correlation vanishes, just as one would expect if the plasma undergoes a crossover from a confined phase to a weakly correlated high temperature phase. The total induced charge, tabulated with the temperature in the figure, provides further information.

Notice that at $0.75T_c$, the induced quark number is considerably different from $-1$. As we noted above, departures from $-1$ are caused by diquark screening. A simple model based on the lowest S-wave baryons and mesons that could be formed in the screening process predicts $Q = -0.81$ at this temperature, considerably closer to $-1$.[52] To get the observed value would therefore require many more baryons than just the low lying S-wave states. Thus there appears to be an anomalously large baryonic component. If this effect is corroborated in further simulations, it might suggest a proliferation of baryons (and antibaryons) in the hadronic plasma as the crossover is approached from below.[53] Many years ago, Hagedorn proposed that an exponentially growing density of hadronic states would lead to a limiting temperature of about $160 \pm 10$ MeV.[54] Of course, we now understand that there is no limiting temperature, since the entropy of hadronic states is limited by the entropy of the underlying quark and gluon degrees of freedom. However, the tendency for the density of hadronic states, particularly baryons, to grow rapidly with mass may lead to an entropy-induced proliferation of these states just below the crossover.

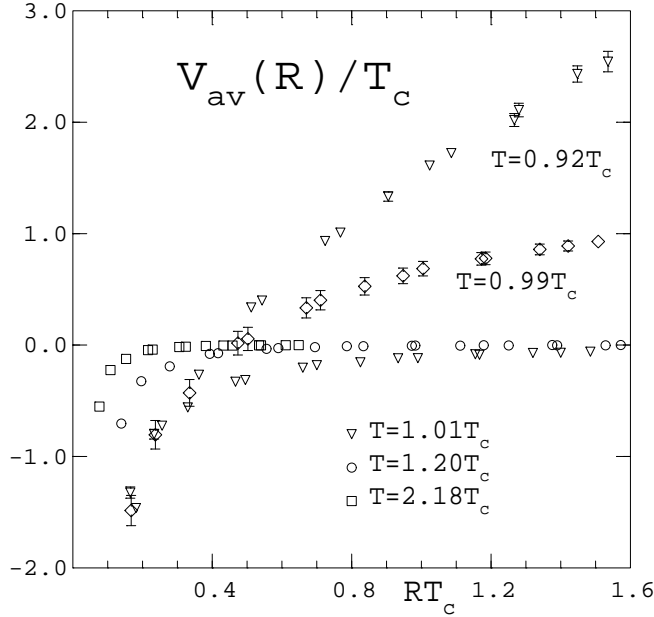

Figure 14: Heavy quark potential in an SU(3) gluon ensemble at various temperatures relative to the deconfinement temperature $T_c$. From Ref. 55.

7.3. *Heavy Quark Potential*

The correlation of two Polyakov loops (31)

$$\exp[-V(r,T)/T] = \langle P(0)P^\dagger(r)/9 \rangle_U \qquad (59)$$

measures the change in free energy $V(r, T)$ of the thermal ensemble brought about by introducing a static quark and antiquark pair. It is a measure of the quark-antiquark potential (free-energy) as a function of separation distance $r$ and temperature $T$. If dynamical quarks are omitted, at low temperature we expect the potential to grow linearly with distance, reflecting confinement. At high temperature deconfinement predicts a potential that becomes asymptotically constant at large $r$. This expectation is borne out in Fig. 14. With light quarks included the heavy quark is screened and we expect the potential to approach a constant at any temperature.

7.4. *Topological Structures*

The effort to develop phenomenological models of QCD, particularly to explain confinement, has frequently employed topological structures, particularly instantons and monopoles. These structures are to be found in lattice configurations. They may play an important role in the high temperature crossover, and they may reveal themselves as nonperturbative contributions to the equation of state. Although our understanding of these structures is still rudimentary, we mention them here for completeness.

7.4.1.*Instantons*

Instantons are typically exposed in lattice configurations after a smoothing or cooling treatment to eliminate high frequency noise.[56] A measure of the importance of instanton configurations is the topological susceptibility. Folklore says that at high temperature instantons should be suppressed, the axial gauge anomaly should disappear, and the $U_A(1)$ chiral symmetry should be restored.[38] Indeed, it was found some time ago in pure SU(N) gauge simulations that the susceptibility, measured after a cooling treatment, falls dramatically at the deconfinement transition.[57] DiGiacomo *et al.* cautioned, however, that the effect depends on way the susceptibility is defined.[58] Thus if we want to correlate a falling topological susceptibility with the suppression of the gauge anomaly and the restoration of $U_A(1)$, it is important that we find a consistent, renormalizable definition.

If instantons play a role in chiral symmetry breaking, then we should see a correlation between topological susceptibility and the chiral condensate. Recent progress in establishing this connection was reported by Teper.[59] In the Stony Brook instanton program, instantons take center stage in controlling the restoration of chiral symmetry and in determining the screening spectrum. In recent work by Ilgenfritz and Shuryak and by Schäfer, Shuryak, and Verbaarschot it is argued that in full QCD, as the chiral phase transition is approached, the light fermion determinant induces an attractive interaction between instantons and anti-instantons.[60] The resulting molecules are predicted to predominate over solitary instantons and anti-instantons. The strong correlation drives the chiral phase transition. Schäfer, Shuryak, and Verbaarschot compute hadronic screening masses in the model and find a spectrum in qualitative agreement with results from lattice simulations. It would be interesting to test their proposals further in lattice simulations.

### 7.4.2. Monopoles

A study of the thermal behavior of QCD may provide insight into the mechanism of confinement. An old 't Hooft—Mandelstam model characterizes the confining QCD vacuum as a dual superconductor, with an electric Meissner effect and a condensate of color magnetic monopoles.[61] Some years ago Schierholz *et al* and Kronfeld *et al* explored the association between confinement and the presence of monopole currents in Yang-Mills theories.[62] Interest has revived recently.[63,64]

To identify monopole currents in a nonabelian gauge theory it is necessary to carry out a U(1) projection of the gauge links. This is done by first fixing a suitable gauge. Popular choices include maximal Abelian gauge and a variety of "unitary" gauges, one of which involves diagonalizing the product of gauge links forming the Polyakov loop, inviting the conclusion that monopoles are responsible for confinement. A "monopole" contribution is then extracted from the resulting U(1) gauge field following the procedure of DeGrand and Toussaint.[65]

In the new approach the string tension and other confinement features are computed using only the monopole contribution to the U(1) field. Good agreement is found with the full string tension computed in the conventional way. In the past year the Kanazawa group has also calculated the projected U(1)-monopole Polyakov-loop expectation value in both SU(2) and SU(3) Yang-Mills theory.[66] The behavior of the monopole-projected Polyakov loop variable imitates the behavior of the conventional Polyakov loop variable. The similarity is seen in a variety of U(1) projection schemes.

Although the results are promising, further work is needed, first, to demonstrate that the suitably defined abelian monopole currents survive the continuum limit[67] and, second, to find a suitable order parameter for monopole condensation.[68]

## 8. Bulk Properties of the Quark Gluon Plasma

Three observables give important information about the characteristics of the quark plasma. The equation of state, giving energy density and pressure as a function of temperature, shows behavior expected in a rapid crossover from a phase dominated by hadrons to a phase dominated by quasi-free quarks and gluons. Departures from an ideal relativistic quark-gluon gas are particularly noticeable in the pressure for $T_c < T < 2T_c$. The baryon susceptibility, which measures the fluctuation in baryon number, also shows behavior typical of deconfinement.

### 8.1. Equation of State

Phenomenological models of the high temperature phase require knowing its equation of state and related bulk thermodynamic quantities, such as the rise in energy density at the crossover, the velocity of sound, and the contribution to the energy density from various species, including strange quarks. Particularly important is the determination of the peak in the specific heat near the phase transition for simulations with quarks. Simulations of the energy density and pressure of the quark plasma are costly because of a low signal-to-noise ratio.

The earliest determinations of the energy density $\epsilon(T, m_q)$ and pressure $p(T, m_q)$ as a function of temperature and quark mass[3] used the basic thermodynamic identities

$$\epsilon(T, m_q)V = \frac{\partial F(V, T, m_q)}{\partial(1/T)} \tag{60}$$

$$p(T, m_q) = \frac{\partial F(V, T, m_q)}{\partial V} \tag{61}$$

In a lattice simulation each such derivative of the free energy involves a separate variation of the spatial and temporal lattice constants $a$ and $a_t$, entailing a concomitant renormalization of the gauge coupling. Some years ago Karsch determined the perturbative asymptotic variation of the gauge coupling with respect to the anisotropy parameter $\xi = a_t/a$.[69] Unfortunately, present simulations are not in the perturbative asymptotic region. Although in principal the Karsch coefficients could be determined nonperturbatively from lattice simulations, this has not yet been done successfully.[26]

Fortunately there is a different nonperturbative route to $\epsilon$ and $p$. The "interaction measure"

$$I = \epsilon - 3p \tag{62}$$

is more easily determined, since it involves an isotropic variation of the lattice constant, requiring only the usual nonperturbative renormalization of the lattice coupling. The pressure, on the other hand, can be determined separately from the free energy density $f = F/V = -p$ by integrating either of two relations[70]

$$\langle \Box \rangle = \frac{\partial F(6/g^2, m_q, V)}{2V \partial(6/g^2)} \tag{63}$$

$$\langle \bar{\psi}\psi \rangle = \frac{\partial F(6/g^2, m_q, V)}{V \partial m_q}. \tag{64}$$

A vacuum subtraction is performed to give the pressure relative to the pressure of the nonperturbative vacuum:

$$\frac{p}{T^4} = 2N_t^4 \int_{\text{cold}}^{6/g^2} d(6/g'^2) \left[ \left\langle \Box(6/g'^2, am_q) \right\rangle_{N_t} - \left\langle \Box(6/g'^2, am_q) \right\rangle_{\text{sym}} \right] \tag{65}$$

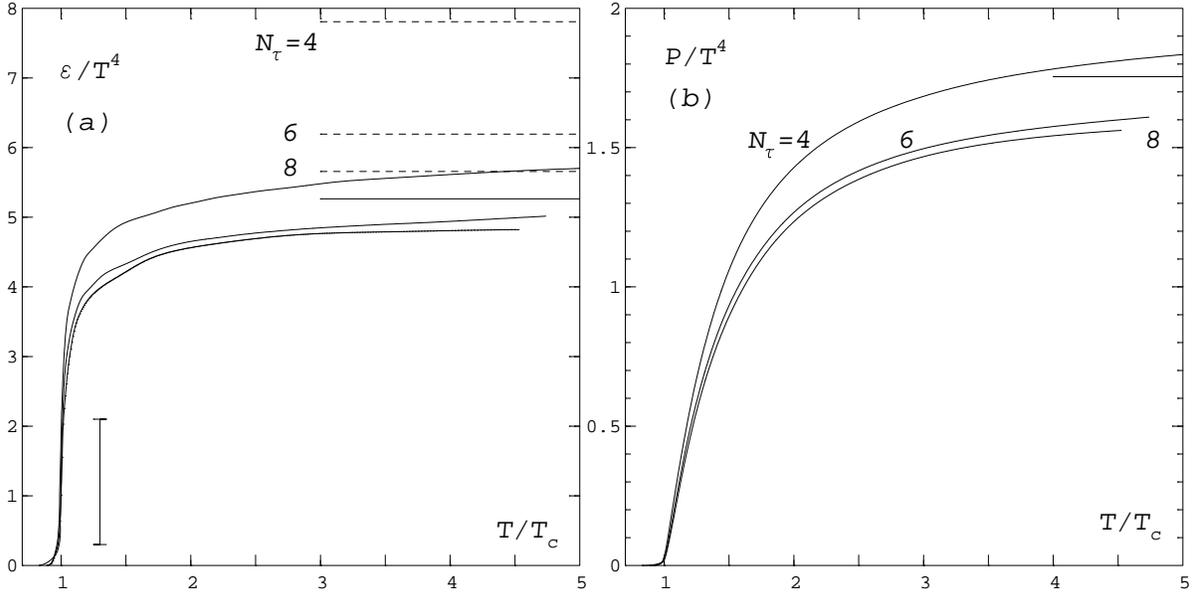

Figure 15: From Ref. 25. (a) Energy density of the SU(3) gluon ensemble *vs.* temperature in units of the critical temperature $T_c$, from simulations on $16^3 \times 4$, $32^3 \times 6$ and $32^3 \times 8$ lattices. The dashed horizontal lines indicate the result expected for an ideal gas on these lattices. The solid horizontal line shows the continuum result. The vertical bar indicates the latent heat discontinuity. (b) Pressure in the SU(3) gluon ensemble *vs.* temperature.

$$\frac{p}{T^4} = N_t^4 \int_{\text{cold}}^{am_q} d(am_q') \left[ \left\langle \bar{\psi}\psi(6/g^2, am_q') \right\rangle_{N_t} - \left\langle \bar{\psi}\psi(6/g^2, am_q') \right\rangle_{\text{sym}} \right] \qquad (66)$$

Of course the latter equation may be used only when dynamical quarks are present. Together with a determination of the interaction measure, this result can then be used to determine the energy density.

These methods have been used in a nonperturbative determination of the energy density and pressure in SU(2) and SU(3) Yang-Mills theory[25,55,71] and in SU(3) with two flavors of dynamical staggered fermions.[26] The pure glue theory can be simulated with higher precision than the theory with quarks included. It tells us in some detail how well QCD resembles a relativistic ideal gas in its bulk thermodynamic behavior. Figure 15 shows the result for the equation of state for the pure glue ensemble in SU(3). The curves are smooth having been derived by integrating a smooth interpolation of the plaquette measurements.

The equation of state for a relativistic ideal gas for SU(N) color with $N_f$ quark flavors is

$$\epsilon = \frac{\pi^2}{15}\left(N^2 - 1 + \frac{7}{4}NN_f\right) T^4 \qquad (67)$$

$$p = \epsilon/3 \qquad (68)$$

The continuum value must be corrected for the lattice discretization. This has been done in Fig. 15. It is clear that to a good approximation, for $T > 2T_c$, as the continuum limit is approached, the equation of state resembles closely that of a relativistic ideal gas. However, the approach to the continuum Stefan-Boltzmann limit is slow. For $T < 2T_c$, the pressure departs significantly from the free gas value and the energy density differs by about 10-20%. One might have expected a strong deviation in the pressure, since mechanical stability requires that it be continuous across a phase transition. It has been suggested that the low momentum gluon modes, those that are most affected by strong interactions, may be responsible for the deviation from ideality.[25] Thus precision studies of the equation of state may give indirect evidence for nonperturbative effects, such as contributions from instanton or monopole configurations in the thermal ensemble.

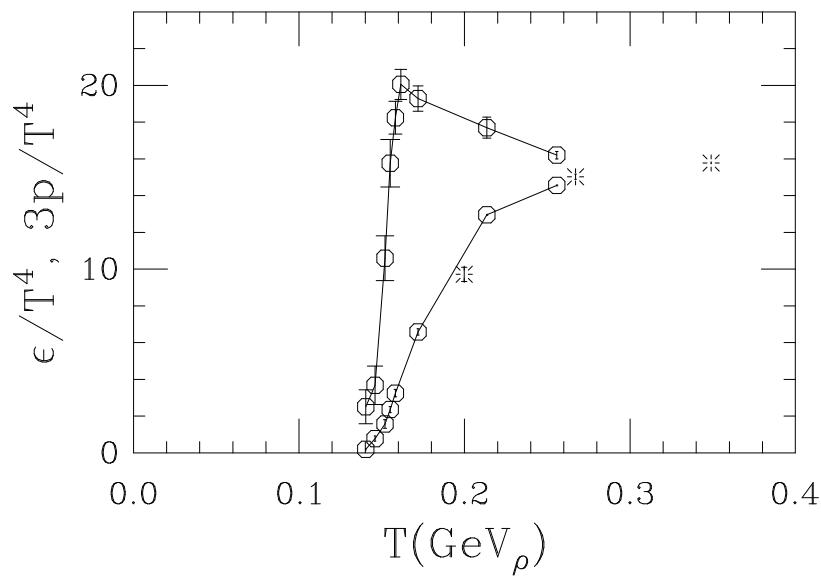

Figure 16: Energy density (upper curve) and three times the pressure (lower curve) vs. temperature (scale based on the rho meson mass) for two light quark flavors ($m_q/T = 0.1$) in the staggered fermion scheme from Ref. 26. The bursts give the pressure extrapolated to zero quark mass

When quarks are included there is no evidence for a bona fide phase transition at nonzero quark mass. Nonetheless, there is a steep rise in the energy density at the temperature associated with the largest slope in the Polyakov loop and $\langle \bar{\psi}\psi \rangle$, as seen in Fig. 16. The transition region is remarkably sharp—of order 20 MeV. In a cooling quark plasma such a strong crossover could cause a momentary slowing in the expansion of the plasma as the quarks and gluons reorganized themselves into more compact hadrons.

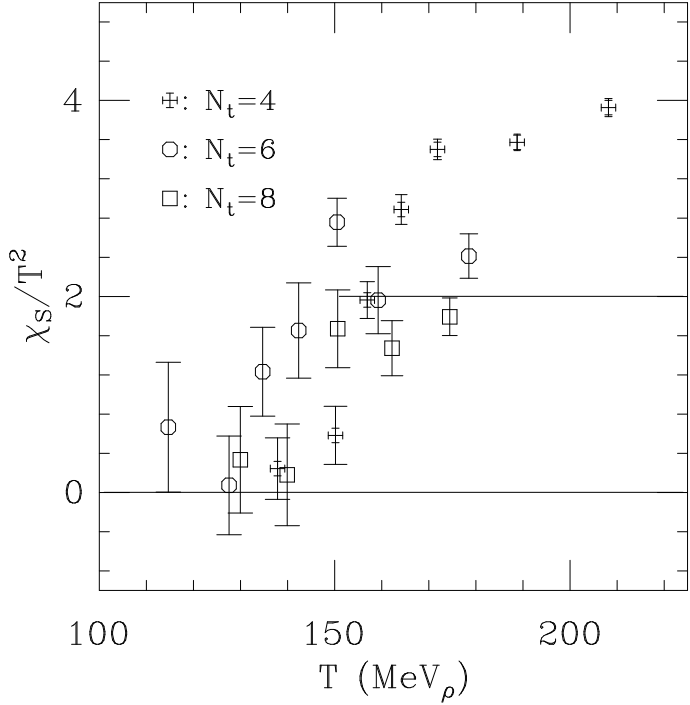

Figure 17: Singlet baryon susceptibility *vs.* temperature for three lattice sizes and for two flavors of staggered fermions from Ref. 24. The horizontal line indicates the continuum ideal gas prediction.

*8.2. Baryon Susceptibility*

Most QCD simulations are done in the grand canonical ensemble (with respect to baryon number) at zero baryon chemical potential. As we have mentioned, technical difficulties have so far thwarted attempts at successful simulations at nonzero chemical potential, at least for credibly large lattice volumes.[13] This situation is unfortunate, since in heavy ion collisions one expects regions of nonzero baryon density. In the grand canonical ensemble at zero chemical potential, the average baryon density is zero. However, there are fluctuations. The baryon susceptibility quantifies these fluctuations:

$$\chi_\mathrm{s} = \left.\frac{\partial B(\mu)}{\partial \mu}\right|_{\mu=0} = \langle B^2 \rangle / V. \tag{69}$$

At low temperature a fluctuation in baryon number requires the addition of a nucleon or antinucleon to the statistical ensemble. Thus we expect fluctuations to be suppressed by a Boltzmann factor $\exp(-m_N/T)$, where $m_N$ is the nucleon mass. If the high temperature phase is characterized by an ideal gas of quarks and gluons, fluctuations are then controlled by the free energy of a quark or antiquark. The ideal gas (continuum) result is

$$\chi_{\text{s}} = N_f T^2 \tag{70}$$

The susceptibility can be measured separately for each quark flavor. For example, with two quark flavors, we may define chemical potentials $\mu_u$ and $\mu_d$ for the up and down quarks, respectively. Then besides the singlet susceptibility we have defined above, we have a nonsinglet susceptibility:

$$\chi_{\text{ns}} = \frac{\partial B}{\partial \mu_u} - \frac{\partial B}{\partial \mu_d} \tag{71}$$

which measures fluctuations in isospin.

The baryon susceptibility has been measured in a few recent simulations.[82] Results are shown in Fig. 17. We see a rapid rise in baryon susceptibility at the crossover, as would be expected from deconfinement. There are significant lattice discretization corrections for small $N_t$ that gradually disappear on larger lattices. After allowing for these corrections, we see that this quantity agrees quite well with the ideal gas prediction in the high temperature phase.

## 9. Correlations and Confinement in the Quark Gluon Plasma

We have already discussed static screening correlators in conjunction with chiral symmetry restoration. In this section we return to hadron correlators and examine the remarkable difference between their imaginary timelike ("temporal") and spacelike behavior. Finally, we discuss the confining features of the high temperature spacelike correlators.

*9.1. Hadron Propagation at Imaginary Time*

The Fourier transform of the screening propagator (48) is the temperature Green's function $G(\omega, \mathbf{p}, T)$:

$$\delta_{ab} G_a(\omega, \mathbf{p}, T) = \int d\tau d^3x \langle H_a(0) H_b(\tau, \mathbf{x}) \rangle \exp(-i\omega\tau - i\mathbf{p} \cdot \mathbf{x}) \tag{72}$$

The static screening correlator (49) is then

$$C_a(z) = \frac{1}{2\pi} \int dp_z \exp(ip_z z) G_a(\omega = 0, p_x = 0, p_y = 0, p_z, T). \tag{73}$$

The screening masses locate the poles in the temperature Green's function in $p_z$, at $im_{an}$. Rather than integrating out the transverse coordinates, more generally, we could consider measuring the screening spectrum at nonzero $\omega$, $p_x$, and $p_y$ to get more information about screening. However, most studies set these variables to zero. Thus the static screening correlator is sensitive to poles in the temperature Green's function at zero frequency and imaginary wavenumber.

An entirely different domain of the temperature Green's function can be reached by measuring the corresponding temporal correlator

$$C_{ab}^t(\tau) = \int d^3x \langle H_a(0) H_b(\tau, \mathbf{x}) \rangle \exp(-i\mathbf{p} \cdot \mathbf{x}), \tag{74}$$

which is related to the temperature Green's function through

$$C_{ab}^t(\tau) = T \sum_{n=0}^{\infty} \exp(i\omega_n \tau) G_{ab}(\omega_n, \mathbf{p} = 0, T) \tag{75}$$

for Matsubara frequency $\omega_n = 2\pi n T$. This temporal correlation function is periodic in the imaginary time variable $\tau$ with period $1/T$. The $\tau$ dependence is controlled by poles (and the continuum states) in $\omega$. The spectral decomposition of the Green's function is given by

$$G_{ab}(\omega_n, \mathbf{p}, T) = \int_{-\infty}^{\infty} \frac{d\omega'}{2\pi} \frac{\rho_{ab}(\omega', \mathbf{p}, T)}{i\omega_n - \omega'}. \tag{76}$$

The spectral density $\rho_{ab}(\omega', \mathbf{p}, T)$ is extremely interesting, since it is controls real time excitations of the plasma. Can it be measured on the lattice? In principle, it can, but in practice, it is extremely difficult. The problem is that in an imaginary time simulation, we measure the Green's function only at the discrete Matsubara frequencies $\omega_n = 2\pi n T$. In fact, because our lattices have a finite $N_t$, we know the Green's function only for a finite set $n = 0, \ldots, N_t - 1$. The mathematical problem, then, is to carry out an analytic continuation from the finite discrete set to the whole complex plane. To make matters worse, the continuation is based on data with statistical uncertainties. To proceed, therefore, we must introduce assumptions about the form of the spectral density. In numerical simulations in condensed matter physics, it is possible to progress by making a "maximum entropy" assumption about the form of the spectral density and by collecting data for $N_t \approx 50$ or more.[72] Present thermodynamic simulations in lattice QCD are very far from approaching this standard.

Despite the difficulties in extracting real-time spectral information from temporal correlators, the Bielefeld group has shown that they can be used to obtain interesting qualitative information about the quark plasma.[73] They compare temporal correlators with what would be expected if the correlations were dominated by free quark and antiquark propagation. (See Fig. 18.) The results show that at low temperature, the correlator has a form that can be fit approximately to a single low-lying mesonic state, whereas at high temperature, the correlator has the form roughly expected of a free quark/antiquark pair.

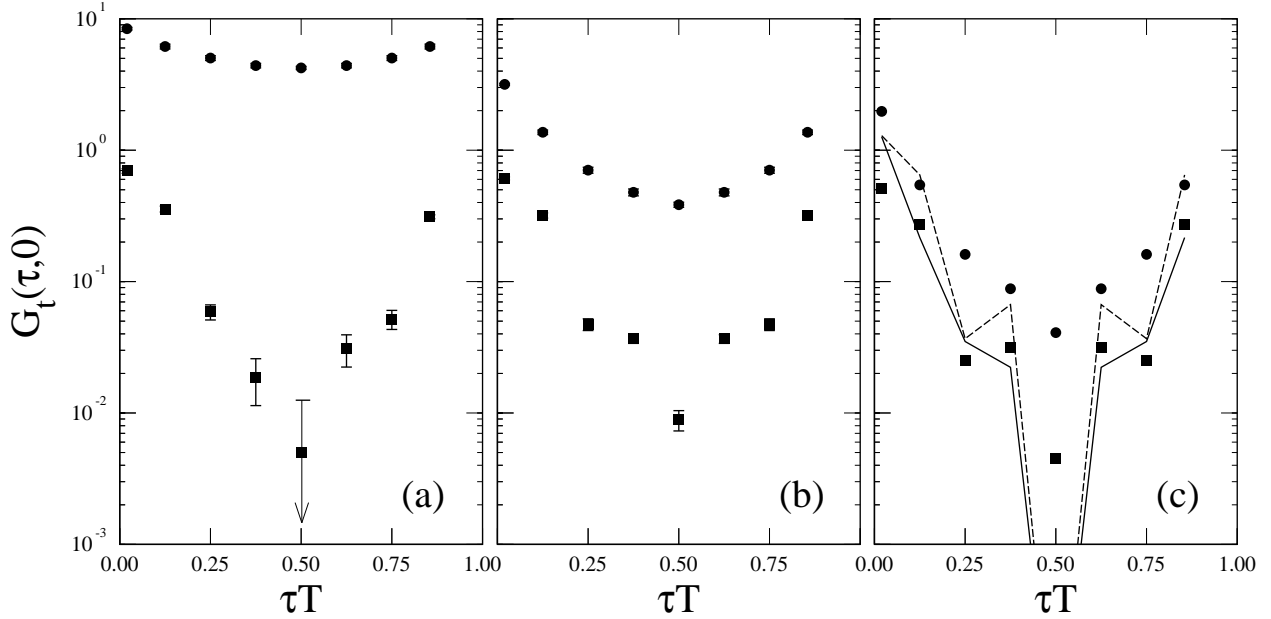

Figure 18: From Ref. 73 The temporal correlators for the staggered fermion pseudoscalar (filled circles) and vector channels (filled squares) at $\beta = 5.1$ (a), 5.3 (b) and 6.5 (c), shown as a function of $\tau T$. In (c) the free quark/antiquark results for the pseudoscalar (solid line) and vector (dashed line) correlators are also shown.

9.2. Screening Wave Functions

We have seen that bulk quantities, such as the energy density, pressure, baryon susceptibility, constituent quark free energy, and temporal correlators all behave at sufficiently high temperature as though the quark-gluon plasma were a relativistic ideal gas. Lest we begin to believe too strongly in an ideal gas description of the plasma, we now revisit the screening states. Recall that the pion screening state is the exchanged object that give rise to the Yukawa interaction between two static nucleons in the plasma. It is predominantly a quark-antiquark state. At low temperature, confinement binds the quark and antiquark to produce the traditional Yukawa pion field. At high temperature, one might expect deconfinement would dissolve the pion, leaving only a $q\bar{q}$ continuum. To explore the structure of this state, one can measure its "wavefunction".

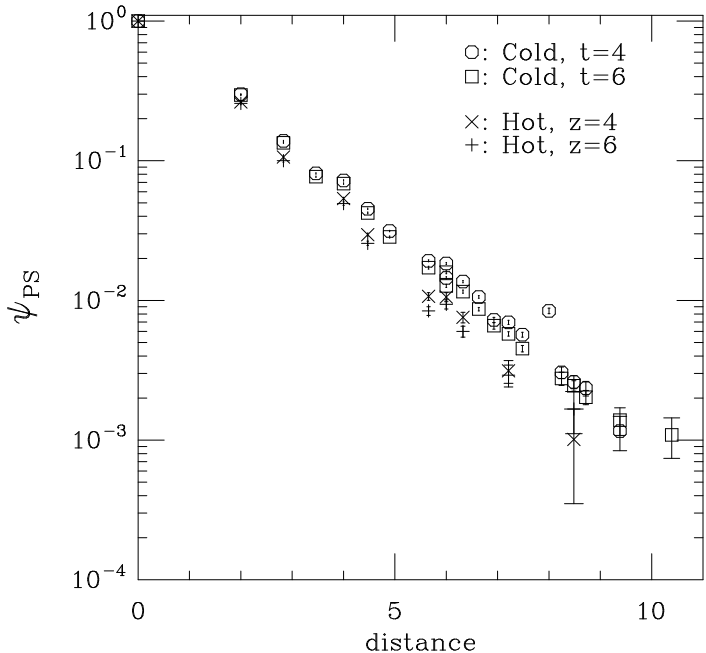

Figure 19: The pion wave function at $T = 0$ and the screening pion correlation function at $T \approx 1.5 \times T_c$ from Ref. 74.

How is a wavefunction measured in a lattice simulation? At zero temperature the wavefunction for the lightest state in a given channel can be obtained by measuring the correlation function

$$\psi(\mathbf{x}) \propto \sum_{\mathbf{y}} \langle \mathcal{O}(t=0) \bar{\psi}(\mathbf{y}, \tau) \Gamma \psi(\mathbf{y} + \mathbf{x}, \tau) \rangle \qquad (77)$$

The operator $\mathcal{O}$ creates a quark and antiquark at $t = 0$ and $\Gamma$ is the appropriate Dirac matrix for the desired meson. The Euclidean time separation $\tau$ should be made large enough so that contributions from excited states die off and the result becomes independent of $\tau$ apart from an overall normalization factor. This correlation function is not gauge invariant, so requires gauge fixing. Coulomb gauge is a popular choice. With $\Gamma = \gamma_5$ we obtain the pion screening state.

The measurement of the wavefunction of the screening state is accomplished in precisely the same manner, after interchanging the roles of the imaginary time coordinate and one of the spatial coordinates–say $z$. Coulomb gauge is then defined with respect to a fixed value of $z$.

A comparison of the wavefunction measured at zero temperature and high temperature is given in Fig. 19.[74] The "zero" temperature lattices of size $16^3 \times 24$ were generated with two flavors of staggered quarks at $6/g^2 = 5.445$ and $m_q = 0.025$. The high temperature wavefunctions were generated on lattices of size $16^2 \times 24 \times 4$ at the same mass, coupling, and flavor number. The high temperature lattice parameters correspond to a temperature of approximately $1.5 T_c$.

The low and high temperature results are strikingly similar, suggesting strong correlations at high temperature. The source operator $\mathcal{O}$ in these simulations was an uncorrelated product of "wall" operators. Thus any correlation must arise from the interaction of the quark and antiquark. The ideal gas result for this observable would give an uncorrelated result independent of separation $r$. Results for other channels, such as the $\rho$ meson and nucleon also show very little change from low to high temperature.[74,75]

*9.3. Dimensional Reduction and Confinement at High Temperature*

The mysterious results for the screening wavefunctions can be explained quite simply from an analysis of the Euclidean path integral.[77] In a Euclidean space all directions are placed on an equal footing. Thus we could just as well call one of the spatial directions, say the $z$ direction, our Euclidean time axis $\tau'$, and call the old periodic Euclidean time direction a new periodic direction $z'$. From this point of view the partition function for nonzero temperature QCD (call it $QCD_t$ for "thermal") is equivalent to the partition function for zero temperature QCD in a three-dimensional space with one of the three dimensions periodic (call it $QCD_c$ for "compact"). Such a variant of QCD is confining, as we shall see. The screening states of $QCD_t$ are just the confined hadronic states of $QCD_c$ and the screening wavefunctions reflect this confinement. As we have observed, the quarks in $QCD_c$ have an effective mass that grows asymptotically with temperature as $\pi T$. At very high temperatures, the periodic coordinate $z'$ becomes so compact, we may speak of a dimensionally reduced theory.[78] In this world all quark states behave like confined heavy quark states. Thus at high temperature it is natural that screening masses are quantized in multiples of $\pi T$ according to their quark content.

If $QCD_c$ is confining, it should be possible to measure the spatial string tension in the theory. This has been done to quite high accuracy by the Bielefeld group. The result is as good a demonstration of confinement in the compactified theory as has been done for the zero temperature four-dimensional theory.[76] A recent high precision result is shown in Fig. 20. For temperatures $T > 2T_c$ the spatial string tension deduced from these measurements is proportional to $g^2(T)T$, as expected from dimensional reduction arguments.

Inasmuch as high temperature meson and baryon screening states are heavy-quark bound states, they lend themselves to a nonrelativistic treatment. The Stony Brook group has explored models of such states.[79]

As we have mentioned, the compactification of the Euclidean time dimension in

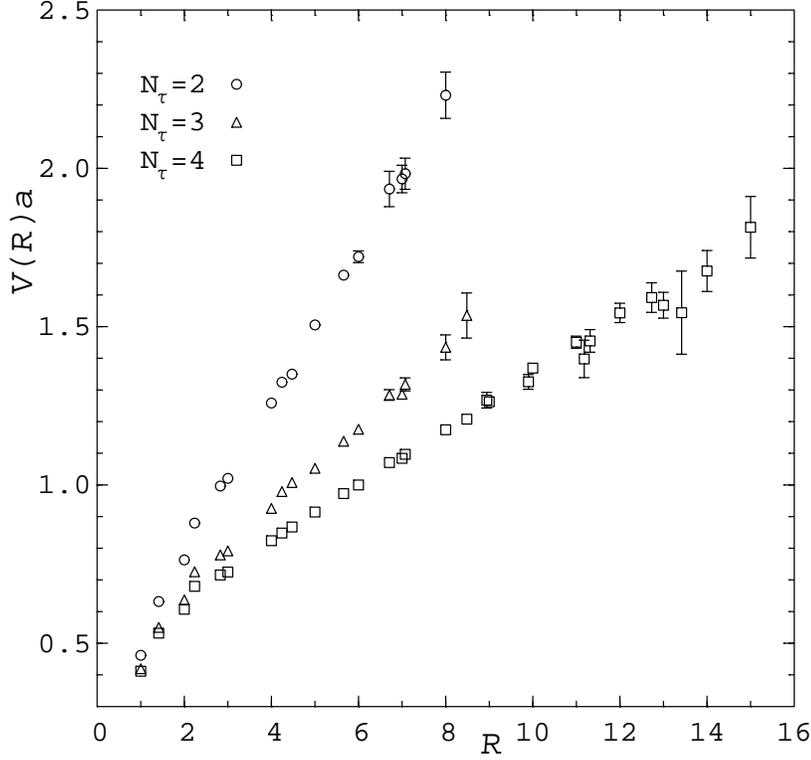

Figure 20: Heavy quark pseudopotential showing confinement in the spatial direction in SU(3) with quarks omitted. Results from Ref. 76 on $32^3 \times N_\tau$ lattices, correspond to three temperatures in the "deconfined" phase with $N_\tau = 2$ (circles), 3 (triangles) and 4 (squares) at $\beta = 6.0$.

QCD results in an effective three-dimensional theory. The four-vector gauge potential $A_\mu^a(x)$ is reinterpreted as an adjoint Higgs scalar $A_0^a(x)$ and a three-vector potential $A_i^a(x)$ for $i = 1, 2, 3$. This observation leads to an intriguing question, important for theoretical insight and model building, at least: Is the resulting theory found in a Higgs phase or a confining phase? The former option entails a spontaneous breaking of the gauge symmetry. In the Higgs phase the infrared behavior of the theory is most likely controlled by 't Hooft-Polyakov monopoles,[80] leading to U(1) confinement. Thus either scenario would result in spatial confinement. In a recent study of SU(2) pure Yang-Mills theory, Kärkkäinen *et al* found strong evidence that the theory chooses the confining phase–that is, the field $A_0(x)$ does not develop a

nonzero vacuum expectation value.[81]

## 10. Conclusions

Through numerical simulations in the staggered fermion scheme, considerable progress has been made in the past several years toward establishing a consistent qualitative picture of the high temperature behavior of QCD. Dynamical quarks make a clear difference in the behavior of the thermal ensemble. Whether there is a phase transition or only a crossover at physical values of the quark masses has not been rigorously established, but it is likely that any first order phase transition has a small latent heat. The crossover temperature is approximately in the range $T_c = 140 - 160$ MeV. The transition from the confining regime to the plasma regime takes place over a relatively narrow range of temperatures (approximately 20 MeV). Viewed over this range, the effective latent heat is large. The plasma is well characterized in bulk as an ideal gas of quarks and gluons for $T \gtrsim 2T_c$, but it retains characteristics of confinement, revealed in long-range screening phenomena, i.e. over distances greater than $O(1/g^2 T)$. Strong deviations from ideality occur below about $2T_c$.

Some gaps remain in our qualitative understanding. The phase structure of QCD with the strange quark included ("2+1 flavors") has not been explored as thoroughly as the two flavor theory. Particularly interesting is the study of critical behavior and soft modes in the multiparameter space of light quark masses. Thermodynamics in the Wilson fermion scheme has yet to establish itself as a contender. Since we need to be confident that the staggered fermion scheme is not misleading us, we must develop better methods for incorporating Wilson quarks. Finally, we need to develop good phenomenological models of the crossover. Topological models offer an intriguing direction. Further work is needed to establish their credibility.

## Acknowledgments

I am indebted to my colleagues for their assistance in preparing this review. In particular, I thank Doug Toussaint, Bob Sugar, Urs Heller, Leo Kärkkäinen, Kari Rummukainen, Tom Blum, Steve Gottlieb, and Krishna Rajagopal. I am also grateful to Frithjof Karsch, Edwin Laermann, and Robert Mawhinney for help with their data. I would like to thank the Institute for Theoretical Physics, University of California at Santa Barbara, where part of this work was carried out. This work was supported in part by the U.S. National Science Foundation under grant NSF-PHY9309458 and by the U.S. Department of Energy under contract DE-FG03-93ER25186.